\documentclass[draft]{agujournal2019}
\usepackage{url} 
\usepackage{lineno}

\usepackage{amssymb}
\usepackage{amsmath}
\usepackage{dsfont}
\usepackage{bm}
\usepackage{float}
\usepackage{ragged2e}
\usepackage{float}
\usepackage{makecell}
\usepackage{multirow}
\justifying

\usepackage{adjustbox}
\usepackage{booktabs}

\renewcommand{\citeNP}{\shortciteNP}

\makeatletter
\def\affiliation#1#2{%
  \vskip-.5\parskip\relax
  {\centering
   {\footnotesize
    \linespread{1.0}\selectfont  
    $^{#1}$#2\par
   }%
  \vskip-.5\parskip
  }%
}
\makeatother

\usepackage{xurl}  

\draftfalse

\journalname{AGU Advances}

\begin{document}

%
%


\def\sysacronym{PROSWIN}
\title{\sysacronym{}: Probabilistic Solar Wind Speed Forecasting Using Deep Distributional Regression From Solar Images}

%
%




\authors{Daniel Collin\affil{1,2}, Yuri Shprits\affil{1,3,4}, Luca Chiarabini\affil{5}, Stefan J. Hofmeister\affil{6}, Nadja Klein\affil{7}, Guillermo Gallego\affil{2,8}}

\affiliation{1}{Space Physics and Space Weather, GFZ Helmholtz Centre for Geosciences, Potsdam, Germany}
\affiliation{2}{Department of Electrical Engineering and Computer Science, Technical University of Berlin,\\
Berlin, Germany}
\affiliation{3}{Institute of Physics and Astronomy, University of Potsdam, Potsdam, Germany}
\affiliation{4}{Department of Earth, Planetary, and Space Sciences, University of California Los Angeles,\\
Los Angeles, USA}
\affiliation{5}{German Aerospace Center, Munich, Germany}
\affiliation{6}{Columbia Astrophysics Laboratory, Columbia University, New York, USA}
\affiliation{7}{Scientific Computing Center, Karlsruhe Institute of Technology, Karlsruhe, Germany}
\affiliation{8}{Einstein Center Digital Future, Berlin, Germany}





\correspondingauthor{Daniel Collin}{collin@gfz.de}




\begin{keypoints}
\item Probabilistic solar wind speed forecasting with well-calibrated uncertainties is possible by using distributional regression.
\item We show advantages of probabilistic over single-value predictions, particularly through risk quantification of fast solar wind conditions. 
\item We provide more informative forecasts than previous models by overcoming a typical trade-off between timeline and HSS peak performance.
\end{keypoints}


\begin{figure}[hb]
    \centering
    \includegraphics[width=1.0\linewidth]{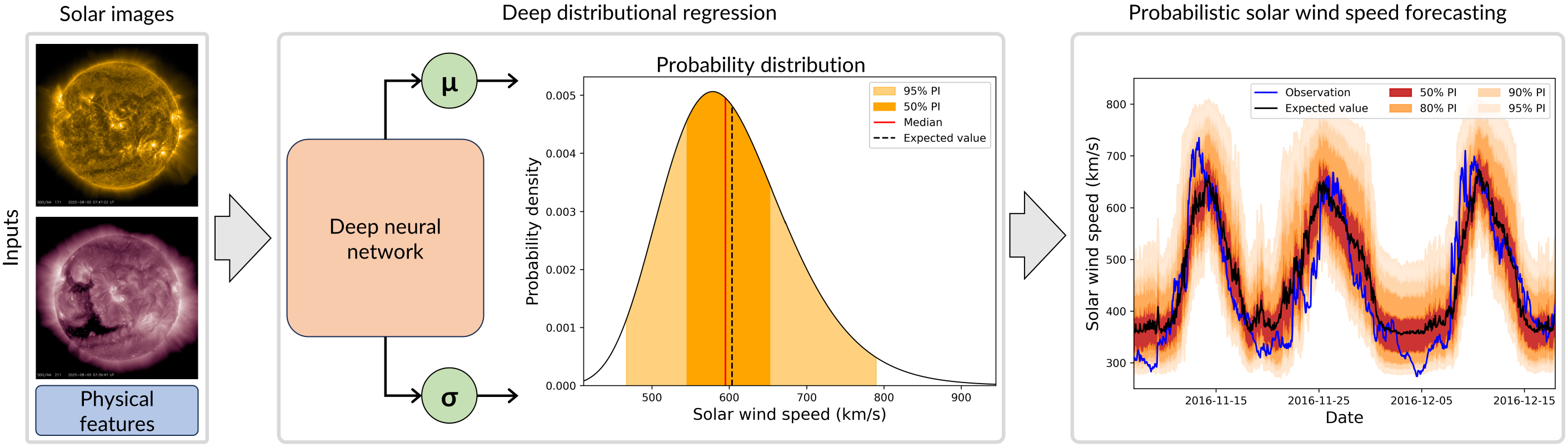}
    \caption{\sysacronym: Solar images and physical features are fed as inputs to a deep neural network, which predicts the distributional parameters of a log-normal distribution for the solar wind speed.
    This enables probabilistic solar wind speed forecasting four days ahead.}
    \label{fig:eyecatcher}
\end{figure}
%
%
%
%


\begin{abstract}

Accurately predicting fast solar wind conditions is challenging, as uncertainties are large and unquantified by traditional single-value prediction models. In particular, the risks of high-speed solar wind streams (HSSs), which can cause damage to technological infrastructure, cannot be reliably assessed without probabilistic forecasts.
We present \sysacronym{}, a probabilistic machine learning model that forecasts the hourly solar wind speed (SWS) at Earth with a four-day lead time. The approach combines solar images and magnetograms using a deep neural network coupled to a distributional regression algorithm. 
Because standard error metrics underweight the relevance of HSS peaks, we further introduce the \emph{prediction score}, a model-selection metric that jointly rewards timeline and HSS peak accuracy.
On 14 years of data, our forecast achieves very well-calibrated uncertainties ($<$1\% average absolute coverage deviation). Using the continuous ranked probability score (CRPS), a metric that assesses distributional accuracy, we obtain a timeline CRPS of 41.0 km/s, an HSS peak CRPS of 45.3 km/s, and a prediction score of 42.3 km/s. We find that the 171~\AA\ channel is an important complement to the typically used 193~\AA\ and 211~\AA\ channels and that the prediction score for model selection improves the applicability of the model. Compared to selected models from the literature, ours is the only one that is accurate for both timeline and HSS peak values, rather than trading one off against the other. 
These results support the advantages of probabilistic over single-value solar wind models. The introduced methods are also transferable to other forecasting problems.
\end{abstract}

\section*{Plain Language Summary}

The solar wind is a stream of charged particles emitted by the Sun that can damage technology in space and on the ground. This occurs especially during periods of fast solar wind, making accurate prediction of the solar wind speed (SWS) essential. However, forecasts of fast solar wind conditions are challenging and often have large errors. To quantify the risks of hazardous conditions, single-value predictions, which give only the most likely outcome, are insufficient. Probabilistic predictions are needed to also capture uncertainty. This study presents a probabilistic forecasting model for the SWS that uses machine learning to learn patterns from solar images. We analyze different image wavelengths and their impact on predictions. Using a new evaluation metric that accounts for the errors of fast solar wind peaks, risk estimation is improved, and the model is better calibrated to operational needs than previous methods. Using 14 years of data, we show that predicted uncertainties are highly reliable, deviating by less than 1\% from observations. Overall, probabilistic forecasts provide more useful information for operators and enable more comprehensive risk assessment than traditional approaches.

\section{Introduction}
\label{sec:introduction}

With the growing number of satellites and space missions, forecasting hazardous space weather effects in real time is becoming increasingly important. 
However, the complexity of the Sun-Earth interactions is not yet fully understood, which limits the accuracy of deterministic single-value prediction models.
Particularly during high-speed solar wind stream (HSS) events, solar wind single-value prediction models frequently make large errors without providing uncertainty information.
This is an important limitation, aggravating decision-making and possibly leading to substantial economic and operational costs.
To ensure that solar wind forecasts are reliable and actionable, they must therefore be provided in a probabilistic way, that is, accompanied by well-calibrated uncertainty estimates \cite{Gneiting14,Camporeale19review}.
While probabilistic forecasting is standard in other fields such as meteorology, most operational space weather models still provide single-value predictions only.

Among the main drivers of hazardous space weather conditions are HSSs and coronal mass ejections (CMEs).
HSSs are plasma streams accelerated along magnetic field lines that are rooted in solar coronal holes and open towards interplanetary space \cite{Krieger73}, whereas CMEs originate from sudden eruptions of solar filaments, ejecting large and fast plasma clouds into the heliosphere \cite{Gosling97}.
During the declining phase and the minimum of the solar cycle, HSSs are the dominant cause of geomagnetic storms, while the frequency of CMEs is related to solar activity and thus increases towards the solar maximum \cite{Richardson00,Tsurutani06}. 
In this study, we focus on the probabilistic prediction of HSSs.

Most operational SWS forecasting models are based on either physical heliospheric simulations (e.g., WSA-ENLIL; \citeNP{Odstrcil03}, EUHFORIA; \citeNP{Pomoell18}) or empirical approaches (e.g., WSA; \citeNP{Arge03}, ESWF; \citeNP{Reiss16,Milosic23}). 
At the same time, machine learning approaches (e.g., \citeNP{Bailey21,Collin25}), especially deep learning from solar images (e.g., \citeNP{Upendran20,Brown22}), where important structures such as coronal holes and active regions can be identified well, show the potential to outperform traditional models \cite{Ahn25}.

However, none of these approaches explicitly predicts forecast distributions, which makes them unable to provide reliable uncertainty estimates. 
Furthermore, they mainly focus on minimizing errors averaged over the whole dataset rather than those of HSS events. 
Due to the strong imbalance between frequent slow speeds and rare HSS peaks, this strategy generally does not lead to actionable forecasts \cite{Owens17prob}, but to a systematic underestimation and large unquantified uncertainties for the most important data points \cite{Shprits19,Collin25,Camporeale25paris}.

There exist several approaches that obtain SWS uncertainties from ensembles (e.g., \citeNP{Bussy-Virat14,Bussy-Virat16,Owens17prob,Issan23,Edward-Inatimi24,Edward-Inatimi26}), whose reliability depends on how the ensemble members are chosen. Ensembles of machine learning models have similarly been used in other applications (e.g., \citeNP{Gu19}).
Probabilistic machine learning models provide an alternative by directly learning the uncertainty from the data as a function of the inputs.
For example, Gaussian Process models and variants have been applied to geomagnetic index prediction (e.g., \citeNP{Gruet18,Chandorkar18,Chakraborty20}). 
Also the direct prediction of distributional parameters for space weather forecasts has been explored in the machine learning context (e.g., \citeNP{Tasistro-Hart21,Bernoux22,Licata22,Hu23,Tahtouh25}). Furthermore, methods have been suggested for learning uncertainty of existing single-value predictions \cite{Camporeale19prob,Camporeale21accrue}.
However, to the best of our knowledge, no probabilistic machine learning model currently exists for the SWS.

In this work, we introduce \sysacronym{}, a Probabilistic Solar Wind speed forecasting approach. It is based on distributional regression \cite{Klein23}, which serves as a general framework for training probabilistic models using arbitrary parametric probability distributions. We model the skewed and heavy-tailed SWS distribution with log-normal distributions, which better describe the uncertainty of the SWS than the commonly assumed Gaussian distributions. 
To learn these distributions directly from solar images and additional physical features, such as the SWS from the previous solar rotation and the state of the solar cycle, we couple distributional regression with deep learning. We optimize the model using a newly proposed metric that balances timeline and HSS peak errors, leading to significant improvements for HSS predictions. 
Then, we investigate the effect of combining different solar image channels and magnetograms, finding that the information content of 193~\AA\ and 211~\AA\ is largely redundant, whereas that of 171~\AA\ is complementary. 
Furthermore, we analyze the reliability of the predicted uncertainties, with a particular focus on HSSs, yielding well-calibrated uncertainties and accurate risk quantification. 
Finally, we demonstrate that our approach outperforms a selection of other solar wind speed models and that the trained models generalize well to new unseen data.
Our contribution also involves the publication of a Python implementation of \sysacronym{} \cite{Collin26code},
as well as the publication of our datasets and trained models \cite{Collin26data}.

The paper is structured as follows: 
Section~\ref{sec:dataset} describes the data and preprocessing. 
Section~\ref{sec:methodology} explains the algorithmic approach, including the distributional regression and deep learning frameworks. 
Section~\ref{sec:evaluation} introduces the evaluation procedure, encompassing a new metric. 
Section~\ref{sec:image_study} studies the impact of different combinations of solar images and magnetograms on the model performance.
Section~\ref{sec:results} analyzes the best-performing model and evaluates the benefits of probabilistic forecasting for SWS prediction. 
Section~\ref{sec:comparison} compares our probabilistic forecast to other methods. 
Section~\ref{sec:unseen} tests the model on unseen data,
Section~\ref{sec:discussion} discusses the main findings and limitations, 
and Section~\ref{sec:conclusion} draws conclusions for future research.

\section{Dataset}
\label{sec:dataset}

Our study combines image data with time series of physical features. 
The image inputs consist of three solar image channels along with magnetograms.
In addition, we incorporate physical features derived from in-situ solar wind measurements and indicators of the solar cycle, resulting in a total of 63 features.
The prediction target is the hourly average of the SWS, measured four days after the input images are recorded. 
The dataset and forecast cadence is one hour, allowing to predict SWS variability on the timescale of HSS onsets and peaks, which can produce sharp increases over a few hours. Furthermore, \citeA{Brown22} show that the performance of their solar image-based SWS model improves with finer training data sampling but with diminishing returns, recovering most of the improvement already at hourly resolution.

Our data period spans the period from June 2010 to June 2026. The last two years of the dataset, from July 2024 onwards, are not taken into account for model development, training, or cross-validation, but are exclusively used for an additional test of the model on previously unseen data (in Section \ref{sec:unseen}).
After preprocessing, this yields 123,129 data points for the main dataset used throughout this study and an additional 17,425 data points for the test on unseen data.
Furthermore, we identify time intervals associated with HSSs and CMEs using the interplanetary coronal mass ejection (ICME) catalog of \citeA{RichardsonCane,RCdata}.
In this section, we describe the dataset in detail, which is also publicly available in \citeA{Collin26data}.

\subsection{Solar Images}
\label{sec:solar_images}

We use solar extreme ultraviolet (EUV) images from the 171, 193 and 211 Å channels of the Atmospheric Imaging Assembly (AIA; \citeNP{Lemen12}), and line-of-sight magnetograms from the Helioseismic and Magnetic Imager (HMI; \citeNP{Scherrer12}), both on board the Solar Dynamics Observatory (SDO).
The 171~Å channel predominantly captures the emission of Fe IX at about 0.8 MK, highlighting mainly the low solar corona. 
The 193~Å channel records Fe XII emission in the corona around 1.6 MK, and the 211~Å channel records Fe XIV emission around 2.0 MK. 
Coronal holes are clearly visible as dark structures in the 193 and 211~Å channels. 
The 171~Å channel provides complementary information related to the expansion factor of coronal holes.
The HMI magnetograms measure the photospheric magnetic field component, providing information on the magnetic polarity distribution and large-scale magnetic structures, such as active regions.
AIA data is available with a temporal cadence of 12 seconds, whereas the magnetograms have a cadence of 12 minutes. Our approach uses both at an hourly cadence. While AIA images and HMI magnetograms would also be available in near-real time with a latency of less than one hour from \url{https://sdo.oma.be/data/aia_quicklook/}, and \url{http://hmi.stanford.edu/magnetic/}, respectively, we use the definitive level~1 science data from the Joint Science Operations Center (JSOC) at \url{http://jsoc.stanford.edu/}.

We preprocess the images, correcting for position, rotation and scaling of the solar disk, light scattering, and instrument degradation, followed by a value normalization and the selection of a central square whose corners are approximately at
the edges of the solar disk. These procedures result in images of $224\times224$ pixels and are adapted from \citeA{Galvez19}, \citeA{Jarolim21}, \citeA{Brown22}, \citeA{Hofmeister24}, and \citeA{Hofmeister25}.
The exact methodology is described in \ref{app:preprocessing}.

\subsection{Solar Wind Measurements}
\label{sec:solar_wind_features}

As SWS data, we use the hourly-averaged plasma bulk speed provided directly by the NASA OMNIWeb database \cite{OMNI}, recorded in-situ at the Lagrange Point 1 (L1). We do not perform any additional averaging ourselves.
In the OMNI database, this dataset is constructed from high-resolution (1–5 minute) upstream measurements that are time-shifted to the Earth’s bow shock using a ballistic convection model, in which each individual observation is shifted according to its own concurrent in-situ plasma speed before hourly averaging. 
This time-shifting procedure aligns the data with measurements of other spacecraft closer to Earth and accounts for spacecraft location, Earth's orbital velocity, and phase-front orientation \cite{King05}.

To account for the periodicity of recurrent coronal holes, we also include as physical inputs the hourly averages of the SWS from 26 to 28 days prior to the forecast time, resulting in 49 features. 
A fraction of values (0.5\%) is missing in the SWS time series. Those gaps are filled by linear interpolation. 
In addition, we include the heliospheric latitude of Earth, i.e., the angle between the solar equatorial plane and the current position of Earth, which is derived from the heliographic inertial coordinates provided in the OMNI\_M data by the NASA COHOWeb. 
This feature captures the latitudinal variation of Earth during its revolution around the Sun, relative to the solar latitudes of the solar wind source region \cite{Hofmeister18}. 
In total, 50 input features are derived from solar wind measurements. 

While OMNI SWS data is typically available with a latency of 20 days, it can at times exceed the 22-day threshold required for our 26–28 day input window. To facilitate a real-time implementation of our approach, OMNI data could be substituted with near-real-time measurements, for example from ACE (\url{https://sohoftp.nascom.nasa.gov/sdb/goes/ace/monthly/}). Although ACE SWS data has a correlation of 0.99 with OMNI SWS data on our data period, minor discrepancies mean real-time performance may not fully match the results reported in this study.
The heliospheric latitude of Earth follows a fixed pattern and can be extrapolated if needed in real-time.

\subsection{Solar Cycle}
\label{sec:solar_cycle_features}

SWS models are known to depend on the solar cycle (e.g., \citeNP{Collin25}). 
Thus, we include as indicators of the solar cycle the twelve most recent values of the monthly sunspot number from the WDC-SILSO at the Royal Observatory of Belgium \cite{SILSO_Sunspot_Number}
and the number of years that have passed since the start of the current cycle (as a coarse indicator of the current position within the solar cycle). 
According to the NOAA Space Weather Prediction Center, solar cycle 24 started in December 2008 and solar cycle 25 started in December 2019. We adopt those solar cycle start dates for our study. 
This yields 13 solar-cycle-related input features. Since the monthly sunspot number is available at the end of each month and we only use the values of past months, all needed features are also available in real-time.

\subsection{High-Speed Streams and Coronal Mass Ejections}
\label{sec:hss_cme}

We focus on forecasting HSSs and therefore need to distinguish between HSSs and CME-related disturbances for the later evaluation of HSS prediction performance. 
However, CMEs are not removed from the dataset. Instead, since our model provides no explicit CME input, we expect CME-driven disturbances to manifest as unexplained variance during training.
To distinguish between both events, we apply the procedure introduced in \citeA{Collin25} to filter all SWS enhancements, i.e., extended periods of elevated SWS, from the solar wind time series. 
Then, we use the ICME list of \citeA{RichardsonCane,RCdata} to classify them.

First, we smooth the SWS time series using a Gaussian filter with a standard deviation of 24 hours and identify local enhancements that exceed 390 km/s and have a peak excess, i.e., vertical distance between its highest point and its base, of at least 35 km/s. 
Next, we define the start and end of each enhancement by the points where the smoothed time series crosses the relative height of 0.4 from the peak's base to its maximum. 
If two events overlap, the longer one is truncated.
Enhancements that overlap with an ICME interval or occur within two days after an ICME are classified as a CME-related disturbance. 
All other events are labeled as HSSs. 
This produces HSS and CME lists with approximate start and end dates, which are later used for model evaluation.

\section{\sysacronym{}}
\label{sec:methodology}

The following sections describe the probabilistic regression model employed, how to leverage predictive performance for the SWS forecasting by combining the model with deep neural networks, and how to perform forecasting. 
We call this approach \emph{\sysacronym{}}.
A Python implementation of the method is available in \citeA{Collin26code}.

\subsection{Distributional Regression}
\label{sec:distributional_regression}

Traditional regression models (e.g., generalized linear or generalized additive models; \citeNP{NelWed1972,HasTib1990}) focus on modeling the conditional mean of a target variable, i.e., the expected value of the target conditional on a set of input features.
This is a relevant limitation, as in many applications, mean-based models can deviate substantially from observations for individual events. 
For the SWS, whose distribution is strongly influenced by fast speeds, this limitation is particularly relevant. 
Focusing on the mean typically leads to severe underestimation of fast events and, consequently, high unquantified uncertainty.
In contrast, distributional regression methods address this limitation by predicting the parameters of a probability distribution for the target variable instead of the single mean value, thereby modeling the entire conditional distribution \cite{Klein23}. 
Here, we introduce the idea of distributional regression applied to SWS forecasting. More details on the implementation and a more general mathematical definition can be found in \ref{app:distributional_regression}.

\begin{figure}[htb]
\centering
\includegraphics[width=1.0\textwidth]{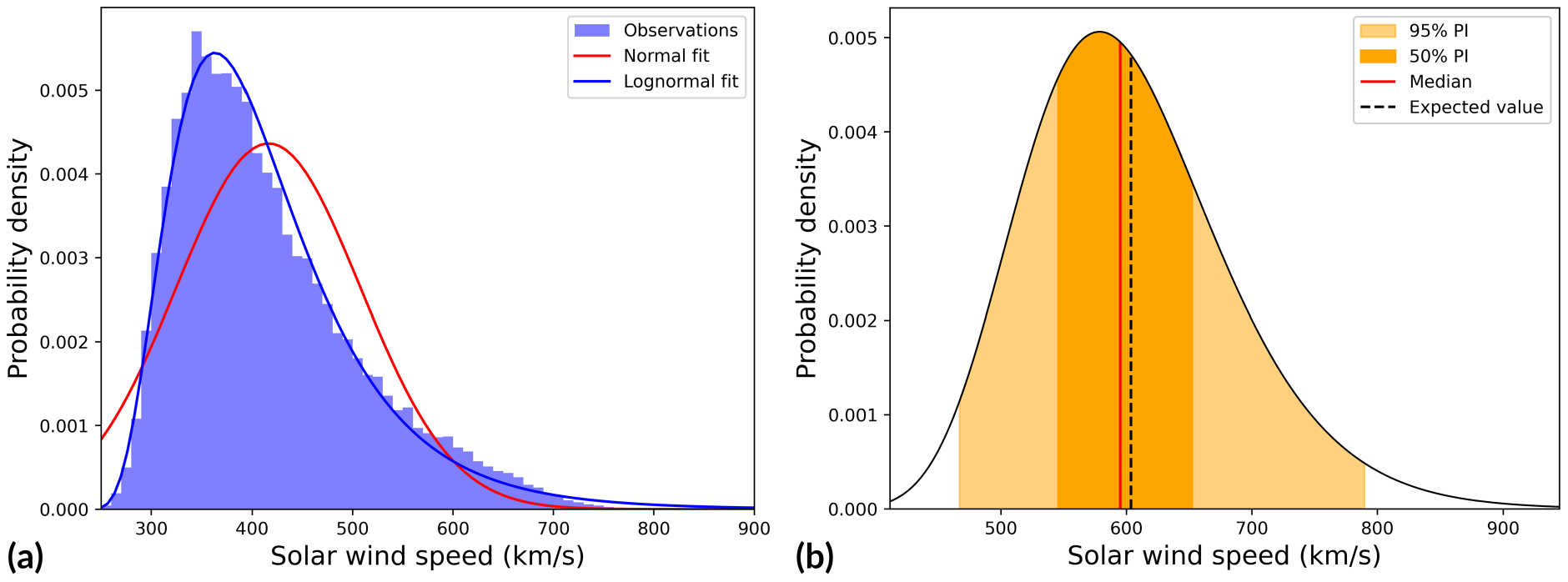}
\caption{(a) Empirical distribution of SWS observations in the dataset, together with the best-fitting normal and log-normal distributions. The log-normal distribution is better suited for the data because of its strict positivity and skewness, particularly modeling the heavy upper tail. (b) Example of a predicted conditional log-normal distribution for a given input. In addition to the expected value, the uncertainty of the prediction is fully quantified.}
\label{fig:dr}
\end{figure}

To apply distributional regression, we first assume that the SWS target variable $V_{\text{sw}}$ is distributed according to a parametric distribution family. 
These are probability distributions which are defined by a set of parameters, e.g., normal, exponential, or uniform distributions, and the distributional parameters typically represent properties such as the location, scale, or shape of the distribution.
Previous mean-based SWS models (e.g., \citeNP{Upendran20,Bailey21,Raju21,Brown22}) are trained using a mean squared error loss, which can be interpreted as implicitly assuming that the target follows a normal distribution with constant variance. 
However, the observed distribution of the SWS is highly skewed with a heavy upper tail. 
This can be seen in Figure~\ref{fig:dr}(a), which shows the empirical distribution of SWS measurements in our dataset together with the best-fitting marginal normal and log-normal distributions, respectively.
While this perspective is marginal only, i.e., not yet accounting for input features, histograms can be helpful for a first visual assessment of which distributional family would be a reasonable choice for the problem. 
Our choice of a log-normal distribution is motivated by its strict positivity, which assigns zero probability to physically impossible negative SWS values, unlike the normal distribution. 
Further, the SWS distribution reflects a quiet baseline of slow solar wind punctuated by episodic high-speed enhancements from HSSs and CMEs, which generates an asymmetry that is captured by the skewness of a log-normal distribution.
Figure~\ref{fig:dr}(a) confirms that the normal distribution underestimates the tail behavior, whereas the log-normal distribution reproduces the observed skewness and large positive deviations.
This characteristic is important for space weather applications, where the most relevant events are located in the upper tail. 
Thus, we assume that $V_{\text{sw}}$ is log-normally distributed with parameters $\mu$ and $\sigma$, where $\mu$ is the mean and $\sigma$ the standard deviation of the log-transformed SWS $\log(V_{\text{sw}})$.

After choosing a suitable distribution family, the objective within the distributional regression framework is to predict conditional distributions for $V_{\text{sw}}$ by modeling the distributional parameters as functions $\mu(x)$ and $\sigma(x)$ of the input features $x$. 
This allows the predicted distributions to change depending on the solar conditions represented by the features.
Figure~\ref{fig:dr}(b) illustrates an example prediction characterized by the probability density function, specifying the likelihood of different SWS values. In contrast, a mean-based single-value prediction model would only provide the expected value. 
More details on the probabilistic prediction are given in Section~\ref{sec:prediction}.

A key flexibility of the approach comes through the freedom of choosing specific forms of the predictor functions $\mu(x)$ and $\sigma(x)$. We learn those using a neural network, which is described in detail in Section~\ref{sec:deep_learning}.
The network's parameters are fitted by minimizing the negative log-likelihood of the observed data.
This corresponds to finding functions for $\mu(x)$ and $\sigma(x)$ that assign high probabilities to the observed SWS. 
This modeling strategy has been popularized as structured additive distributional regression \cite{KleKneLanSoh2015} or Generalized Additive Models for Location, Scale and Shape (GAMLSS; \citeNP{RigSta2005}).

In summary, we probabilistically model the SWS as $V_{\text{sw}} \sim \mathrm{Lognormal}\big(\mu(x), \sigma(x)\big)$, where $\mu(x)$ and $\sigma(x)$ are represented by neural networks trained using a negative log-likelihood loss. For any model input $x$, this formulation then yields a full conditional distribution of the SWS.

\subsection{Deep Learning of Predictors}
\label{sec:deep_learning}

\begin{figure}[htb]
\centering
\includegraphics[width=\linewidth]{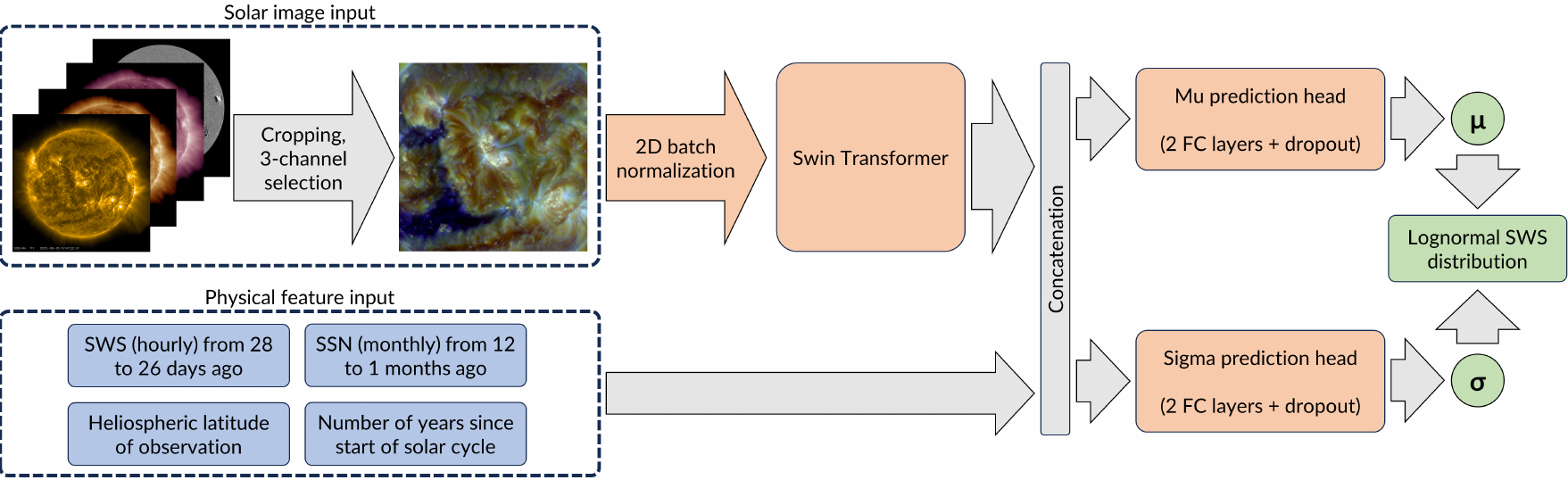}
\caption{Deep learning model pipeline. Information from images and physical features is combined and used in separate prediction heads to model the distributional parameters of conditional log-normal SWS distributions. Red boxes: trainable functions; blue boxes: input data; green boxes: output data.}
\label{fig:model}
\end{figure}

Next, we introduce the neural network that is used to predict the distributional parameters $\mu$ and $\sigma$.
The model architecture is illustrated in Figure~\ref{fig:model}. 
It is motivated by the findings of \citeA{Brown22}, who showed that attention-based neural networks outperform convolutional neural networks for SWS prediction from solar images. 
We therefore adopt the usage of a Swin Transformer \cite{Liu21} as image encoder. 
It has also been shown to be effective in encoding expansion factor and angular distance maps in the WSA+ approach \cite{Mayank25}. 
A Swin Transformer is a special type of Vision Transformer, which partitions an image into patches that are flattened, linearly projected, and processed by self-attention layers. 
While a standard Vision Transformer employs fixed patches, a Swin Transformer uses smaller windows that are shifted between layers, so patches can interact across window boundaries, and a hierarchical structure progressively merges patches.
This allows the model to capture long-range spatial dependencies while maintaining computational efficiency.

As the Swin Transformer supports an input of size $224 \times 224 \times 3$, we select up to three SDO channels and stack them to obtain a three-channel image of $224^2$ pixels. 
If fewer than three channels are used, the selected channels are replicated until a 3-channel image is obtained. 
Because the different channels interact with each other through the attention mechanism, the combination of multiple SDO channels can be exploited.

We initialize the Swin Transformer with weights pretrained on the ImageNet-1k dataset, a large benchmark dataset widely used to learn robust feature representations of images \cite{Deng09}. 
We use the tiny version of the Swin Transformer, as larger variants only yielded minor performance improvements while substantially increasing training time in our experiments. 
Before feeding images into the encoder, we apply a 2D batch normalization layer, which normalizes activations across the mini-batch, reducing internal covariate shift, accelerating convergence, and reducing sensitivity to weight initialization \cite{Ioffe15}.

The output of the image encoder is a low-dimensional vector representation, which is concatenated with the physical feature vector. 
Because we aim to predict both $\mu$ and $\sigma$, the model contains two prediction heads with identical structure. 
Each head consists of two fully connected layers, i.e., one hidden layer, with dropout applied to the hidden layer. 
Dropout randomly removes a fraction of neurons during training to reduce overfitting and smooth the output \cite{Srivastava14}. 
A LeakyReLU activation function is used in both heads, which allows for a small, non-zero gradient when a neuron is not active, and thereby is more robust during optimization \cite{Maas13}.
The combined feature vector is passed to the prediction heads, producing estimates of $\mu$ and $\sigma$, which fully specify the conditional distribution. 
The number of hidden neurons and the dropout rate of both heads are determined by a hyperparameter optimization procedure described in Section~\ref{sec:hpo}. 
The neural network has around 27.7 million trainable parameters, depending on the exact hidden layer size. 
However, 27,519,354 of those are pretrained parameters of the Swin Transformer and only need to be finetuned, whereas around 200,000 parameters need to be trained from scratch.

To improve the numerical stability of the training, we apply some data transformations and train the model in two stages: first the image encoder and the $\mu$-head, then the $\sigma$-head, exploiting the fact that $\mu$ and $\sigma$ decouple for normal distributions during training on log-transformed targets. 
Since the expected value of a log-normal distribution depends on both parameters jointly, this sequential procedure does not imply statistical independence of $\mu$ and $\sigma$. In the final log-normal output, $\mu$ and $\sigma$ are mathematically covariant.
We therefore apply an explicit correction to $\mu$ after fitting $\sigma$, adjusting it so that the expected value of the resulting log-normal prediction matches the value learned by the $\mu$-head in the first training stage. 
This correction enforces the physical integrity of the forecast despite the sequential training procedure.
In addition, we assign higher training weights to HSS events to account for the class imbalance between fast and average solar wind conditions, and a regularization of the SWS uncertainty to enforce physically realistic boundaries.
All technical details on the training process as well as the derivation of the correction and regularization are given in \ref{app:training} and \ref{app:regularization}, respectively.

\subsection{Probabilistic Prediction}
\label{sec:prediction}

After training, the neural network can be used for inference of arbitrary inputs of solar images and physical input features at a particular time point, giving as output distributional parameter estimates $\hat\mu$ and $\hat\sigma$ and thereby predicting a log-normal SWS distribution. 
This probability distribution is characterized by the probability density function (PDF), which represents the relative probabilities of $V_{\text{sw}}$ attaining a specific value or falling within a certain range.
The PDF can be used to derive any quantity of interest for the predicted distribution, for instance, the expected value, which is the typical output of a single-value prediction model, or the cumulative distribution function (CDF), which specifies the probability that $V_{\text{sw}}$ stays below a certain threshold. 
By subtracting the CDF from 1, we further get the probability that $V_{\text{sw}}$ exceeds that threshold.
Finally, we get the quantile function, which is the inverse of the CDF.
It can be used to construct prediction intervals (PIs) that contain $V_{\text{sw}}$ with a specified probability.
All additional technical details about the described quantities are given in \ref{app:pred}.

\section{Evaluation Procedure}
\label{sec:evaluation}

\subsection{Metrics}
\label{sec:metrics}

We differentiate between model performance averaged over the full dataset and the performance for HSSs and their peak values. 
In the first case, metrics are applied to the aggregated time series and are therefore called \emph{timeline} (i.e., full-series) metrics. Metrics that are applied to the subset of observed HSS peaks and their associated predicted HSS peaks are called \emph{HSS peak value} metrics.
Additionally, we combine both performance aspects in a single new metric, which we call \emph{prediction score}. 
In the following, we first introduce the timeline metrics we use, then explain the HSS prediction evaluation, and finally introduce the prediction score.

\subsubsection{Timeline Metrics}

\begin{figure}[htbp]
\centering
\includegraphics[width=0.5\textwidth]{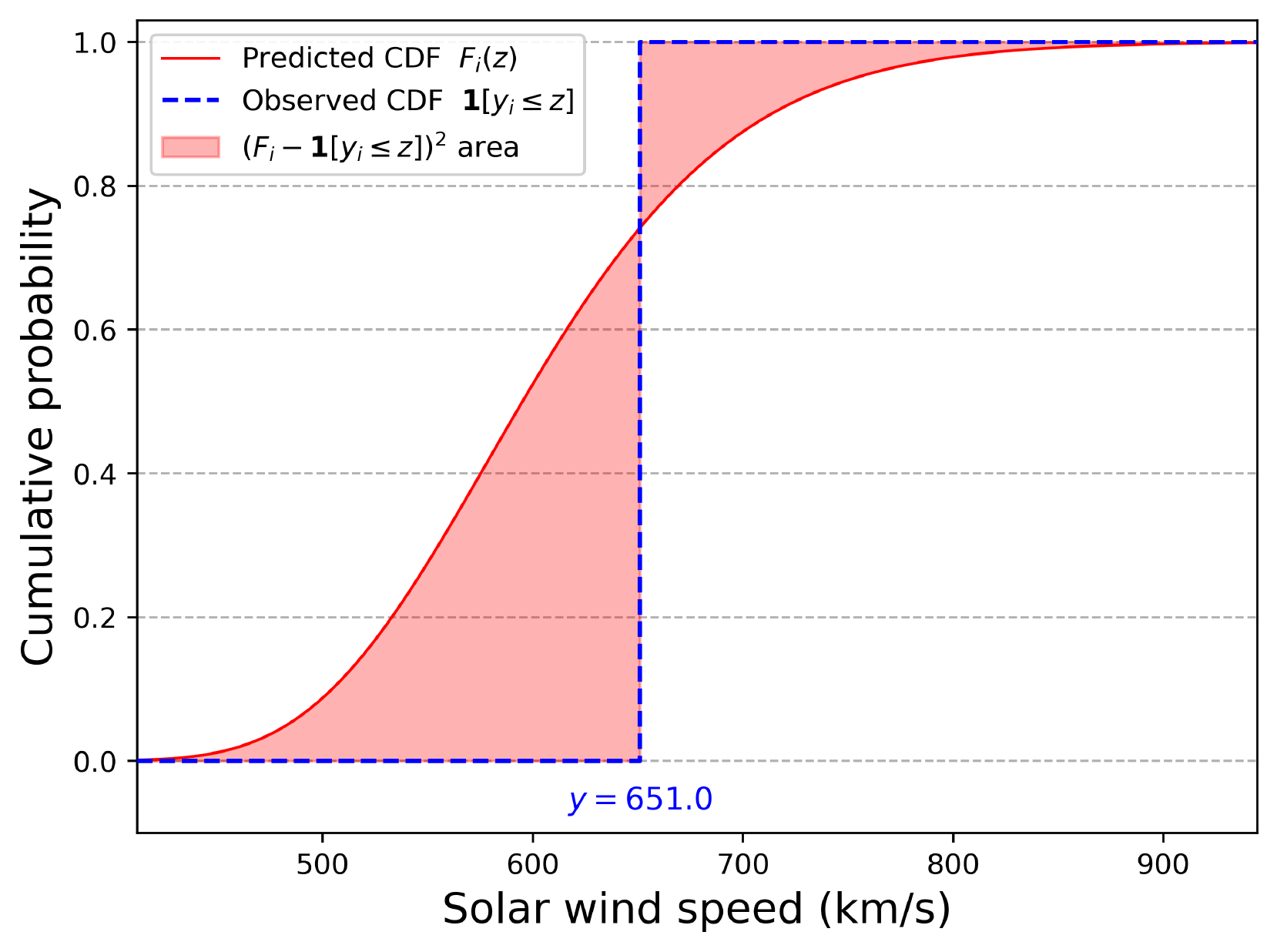}
\caption{Illustration of CRPS computation. The difference between predicted and observed CDF is integrated, such that both sharpness and calibration of predictions are evaluated.}
\label{fig:crps}
\end{figure}

To evaluate the predicted distributions, we use the continuous ranked probability score (CRPS; \citeNP{Gneiting07}), defined as
\begin{equation*}
    \mathrm{CRPS}=\frac{1}{n}\sum_{i=1}^n\int_{-\infty}^{\infty}\big(F_i(z\mid\mu_i,\sigma_i)-\mathds{1}\{y_i \le z\}\big)^2\,dz.
\end{equation*}
It measures the difference between the CDFs $F_i(z\mid \mu_i,\sigma_i)$ of the predicted SWS distributions, based on the parameter estimations $\mu_i$ and $\sigma_i$, and the empirical CDFs $\mathds{1}\{y_i \le z\}$ of the actual SWS observations, which are in fact step functions, because the observed values $y_i$ are attained with 100\% probability.
These step functions are defined using the indicator function $\mathds{1}$.
As those CDFs are probability distributions defined for all real numbers, we compute their difference by integrating over the real line, with $z$ being the variable of integration which represents all possible SWS values. 
A visualization of the computation is shown in Figure~\ref{fig:crps}.
Then, we compute the average CRPS value over all $n$ data points. 
The CRPS rewards sharp and well-calibrated distributions by penalizing forecasts that are both inaccurate (far from the truth) and uncertain (too spread out). 
Lower values indicate better performance. 

We further use the Brier score (BS; \citeNP{Brier50}) to evaluate predicted probabilities $p_i$ of certain events,
defined as
\begin{equation*}
    \mathrm{BS}=\frac{1}{n}\sum_{i=1}^n \left(p_i-o_i\right)^2.
\end{equation*}
Those events can be, for example, the SWS exceeding a specific threshold or the occurrence of an HSS. 
The predicted probabilities are compared to the observations $o_i$, which are 1 if the event occurred, and 0 if not.
The BS can take on values between 0, meaning a perfect forecast, and 1, the worst.
Both CRPS and BS are commonly used for probabilistic prediction models (e.g., \citeNP{Bernoux22,Klein23ts,Edward-Inatimi24,Tahtouh25}).

However, for rare events, e.g., the SWS exceeding 700 km/s, the BS can be positively biased, because even underperforming methods can have a good BS. 
For example, predicting a constant 0\% probability is the perfect prediction for most time points due to the rarity of the events, but clearly has no forecast skill.
Therefore, the Brier Skill Score (BSS) is often considered as an alternative. 
It assesses forecast skill relative to a reference model, which we choose as the naive climatological model that always predicts the observed frequency of the event in the dataset. 
That yields the definition
\begin{equation*}
    \mathrm{BSS}=1-\frac{\mathrm{BS}}{\mathrm{BS}_{\text{ref}}},
    \quad \mathrm{BS}_\text{ref} = \frac{1}{n}\sum_{i=1}^n \left(p_\text{ref}-o_i\right)^2, 
    \quad p_\text{ref} = \frac{1}{n}\sum_{i=1}^no_i.
\end{equation*}
The reference model represents a model without predictive skill, which one seeks to improve on. 
The value $\mathrm{BSS}=1$ represents the best possible score. 
$\mathrm{BSS}>0$ means that the model has an advantage over the reference model, 
$\mathrm{BSS}<0$ means that the performance is worse than the reference model.

In addition, we evaluate the accuracy of the expected SWS values by computing the root mean square error (RMSE) between the observations and the expected values of the predicted distributions.

For the CRPS and RMSE metrics, we report the error uncertainty using 95\% confidence intervals via a moving block bootstrap \cite{Kuensch89} with 1000 replicates, summarized as a symmetric $\pm$ half-width around the single-value estimate. We use a block length of 27~days for timeline metrics, to account for the autocorrelation of the solar wind, and i.i.d. resampling for peak metrics.
We do not report confidence intervals for the event detection metrics or the prediction score, as these are not directly computed from the prediction errors and require more complex resampling schemes.

\subsubsection{HSS Event and Peak Metrics}

To assess the performance of the model for HSSs, we use an event-based approach. 
We identify SWS enhancements in the time series of predictions of the expected SWS values, following the procedure in Section~\ref{sec:hss_cme}, and match them to observed enhancements as described in detail in \citeA{Collin25}, and shortly summarized as follows.
Two enhancements are associated if they occur within three days of each other. 
If several candidate matches exist, we use the temporally closest pair. 
Predicted enhancements associated with observed CME disturbances are removed. 
All other associations are marked as a hit (correct HSS prediction). 
Predicted enhancements that are not matched to an observation are labeled as false alarms (HSS predicted but not observed), and observed enhancements that are not matched to a prediction are labeled as misses (HSS observed but not predicted). 
We count the number of hits as true positives (TP), misses as false negatives (FN), and false alarms as false positives (FP). 
Based on these counts, we compute the verification measures: 
probability of detection (POD), false alarm ratio (FAR) and threat score (TS) \cite{Woodcock76},
\begin{equation*}
    \text{POD}=\frac{\rm TP}{\rm TP + \rm FN},\qquad
    \text{FAR}=\frac{\rm FP}{\rm TP + \rm FP},\qquad
    \text{TS}=\frac{\rm TP}{\rm TP + \rm FP + \rm FN}.
\end{equation*}
The POD quantifies the detection ratio among all observed HSSs, 
the FAR quantifies the ratio of false alarms among all predicted HSSs, 
and TS balances both aspects, rewarding high detection rates only if they are not at the cost of many false alarms. 
All three metrics range between 0 and 1, where 1 is the best outcome for POD and TS, and 0 is best for FAR.

To further evaluate the accuracy of HSS peaks, we also compute the CRPS and RMSE between the observed and predicted HSS peak velocities.

\subsubsection{Prediction Score}

Previous work has shown a systematic trade-off between timeline performance and HSS peak performance in solar wind prediction models (e.g., \citeNP{Shprits19,Collin25}). 
To account for this trade-off, we introduce the prediction score (PS), a composite metric that combines the timeline and HSS peak value errors into a single criterion. 
The prediction score is based on a weighted harmonic mean of the two error components, which favors models achieving good performance across both aspects over optimizing a single error. 
The relative importance of the two components can be adjusted through weighting factors depending on the objective of the model and the characteristics of the dataset.
We define the prediction score as
\begin{equation*}
    \text{PS}_w(e_\text{t},e_\text{p})=\frac{(w_\text{t} + w_\text{p})}{\frac{w_\text{t}}{e_\text{t}}+\frac{w_\text{p}}{e_\text{p}}},
\end{equation*}
where $e_\text{t}$ and $e_\text{p}$ are the timeline and peak errors, respectively.
We assign double weight to the timeline component, i.e., $w_\text{t}=2$ and $w_\text{p}=1$, to counteract an imbalance in the number of underlying data points: The timeline error is computed from all 123,129 hourly predictions, while the HSS peak error is computed from only 284 observed peaks. 
Under equal weighting, each peak prediction would have a disproportionate influence on the score, incentivizing the model to trade general accuracy for isolated peak improvements, e.g., via noisier predictions leading to unphysical spikes in the predicted time series. 
We found empirically that a 2:1 weighting yields the best balance between timeline and HSS peak errors in our experiments.

With our concrete choice of weights, the prediction score simplifies to
\begin{equation*}
    \text{PS}(e_\text{t},e_\text{p})=\frac{3e_\text{t} e_\text{p}}{{e_\text{t}}+2{e_\text{p}}}.
\end{equation*}

\subsection{Cross-Validation}
\label{sec:cv}

We evaluate our model using 5-fold cross-validation (CV). First, the dataset is divided into five subsets. Then, for each fold, three subsets are used as training data to fit the model, one as validation data to monitor overfitting, and one as test data to evaluate the model on unseen data. The overall model performance is assessed by computing the evaluation metrics on the aggregated set of all test data predictions.
Each prediction contributing to this aggregated evaluation set is genuinely out-of-sample, as it comes from the one fold in which that point was held out as test data, and its model never observed it during training or validation. 
All results throughout this manuscript are based on these concatenated out-of-fold predictions from a single 5-fold CV run.

\begin{figure}[htbp]
\centering
\includegraphics[width=0.6\textwidth]{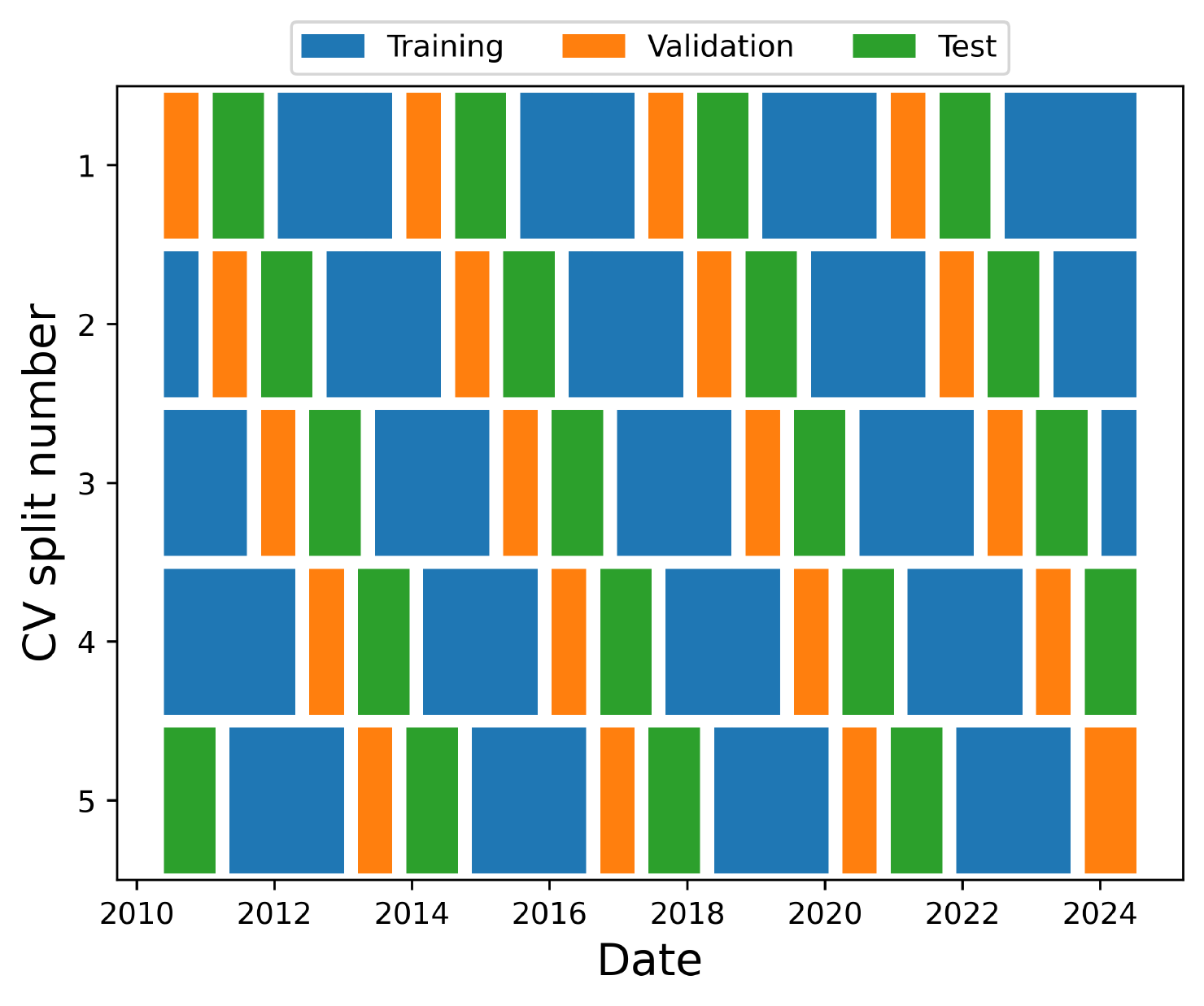}
\caption{Cross-validation data splits of our dataset. Training and evaluation data is distributed over the solar cycle and 90 days are discarded between chunks to avoid data leakage.}
\label{fig:cv}
\end{figure}

The SWS time series exhibits an auto-correlation for up to 4 days and recurring auto-correlation peaks every multiple of 27 days, caused by long-lasting coronal holes that become geoeffective with each solar rotation. 
To avoid data leakage caused by recurring coronal holes appearing in both training and test data, we discard 90 days of data between training, validation, and test datasets. 
Furthermore, the state of the solar corona and correspondingly the solar wind parameters vary over the solar cycle \cite{Richardson00,Tsurutani06}. 
Therefore, we ensure that data subsets are distributed across the solar cycle. To form the five subsets, we divide the dataset into 20 continuous chunks of approximately nine months each, and assign to each subset four of those chunks, while ensuring that each data point appears exactly once as test data. 
The resulting CV splits can be seen in Figure~\ref{fig:cv}.

\subsection{Hyperparameter Optimization}
\label{sec:hpo}

To identify optimal hyperparameters, we run a hyperparameter optimization with the tree-structured Parzen estimator (TPE), a sequential, greedy algorithm based on the expected improvement criterion \cite{Bergstra11,Bergstra13}. For the $\mu$ prediction head, we optimize the expected value prediction of the model, consequently minimizing the prediction score derived from the RMSEs, while for the $\sigma$ prediction head, we optimize the uncertainty prediction of the model by minimizing the prediction score derived from the CRPSs. For each head, we tune the learning rate, batch size, hidden layer size, and dropout rate. We perform 100 TPE iterations for the $\mu$ head and 150 iterations for the $\sigma$ head. 
The resulting best-performing hyperparameters are reported in \ref{app:hyperparameters}.

Since each hyperparameter trial is evaluated on the same cross-validation splits with hyperparameters selected based on the aggregated cross-validation prediction score, this introduces a well-known selection bias \cite{Cawley10}, which can lead to overoptimistic generalization estimates. 
A fully nested cross-validation, with an outer loop held out of the optimization entirely, would be the standard solution, but our current optimization already requires several weeks of compute time to perform all experiments, making an additional outer loop computationally prohibitive. 
We instead mitigate this risk through high dropout rates and a modest optimization budget, and further evaluate our models on a new unseen data period excluded from this entire procedure (Section~\ref{sec:unseen}) to assess whether generalization is compromised.

\section{Study of Solar Image Channels and Magnetograms}
\label{sec:image_study}

\begin{figure}[ht]
\centering
\includegraphics[width=1.0\textwidth]{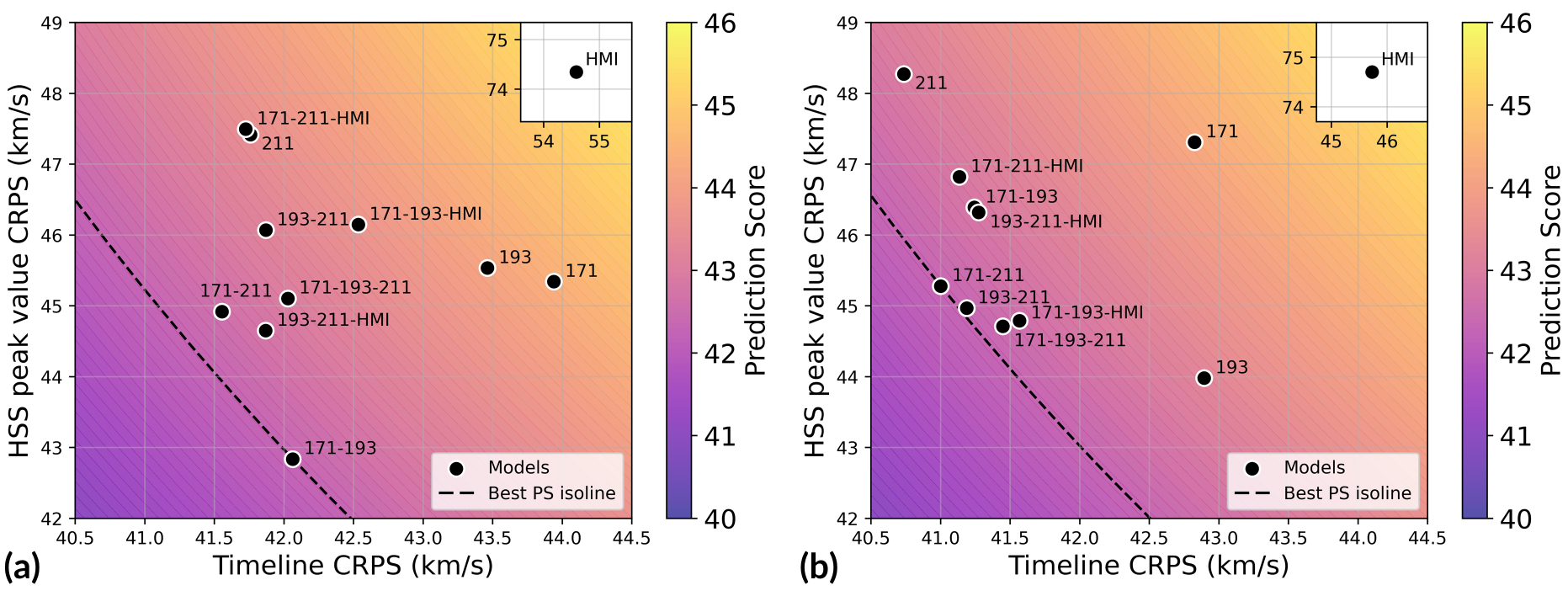}
\caption{Timeline CRPS vs. HSS peak value CRPS of models for different combinations of solar image and magnetogram inputs for the period 2010-2024. 
The trade-off between both errors is visualized by the prediction score, which combines both metrics. The dashed curve shows the isoline of the best model's prediction score. The point labels show the used inputs. Panel (a) shows models that only use images as input (see Table~\ref{tab:nophy}). Panel (b) shows models that use images and the available physical input features as input (see Table~\ref{tab:full_model}).}
\label{fig:channel_study}
\end{figure}

In this section, we study which SDO AIA solar image channels, together with HMI magnetograms, are most suitable for SWS forecasting by evaluating a range of input combinations. First, we remove the set of physical input features, which informs the model about the previous solar wind conditions and the state of the solar cycle, and compare models that only use image data as input. 
These results are investigated in Section~\ref{sec:image-only}. 
Then, we use the full set of available input features, which is analyzed in Section~\ref{sec:full_model}. 
For each model, we perform the hyperparameter optimization described in the previous section. 
The optimized hyperparameters are reported in \ref{app:hyperparameters}. 
Figure~\ref{fig:channel_study} summarizes the CRPS prediction scores, visualizing the trade-off between timeline and HSS peak value errors for both model classes. In the following, we analyze the results.

\subsection{Image-Only Model}
\label{sec:image-only}

Table~\ref{tab:nophy} shows the metrics for the models that use only solar images as input. 
Figure~\ref{fig:channel_study}(a) visualizes the CRPS prediction scores. The prediction score isolines are approximately parallel to the highlighted isoline (dashed curve) of the best model.

\begin{table}[th]
\caption{\label{tab:nophy}Evaluation metrics of image-only models for different combinations of solar image and magnetogram inputs. CRPS and RMSE values are in km/s. The 95\% confidence interval half-width is shown in the second row for each model, where available. Bold numbers indicate the best value per column.}
\centering
\adjustbox{max width=\linewidth}{%
\begin{tabular}{lccccccccc}
\toprule
& \multicolumn{2}{c}{Timeline} & \multicolumn{2}{c}{HSS peak values} & \multicolumn{3}{c}{HSS events} & \multicolumn{2}{c}{Prediction Score} \\
\cmidrule(l{2mm}r{2mm}){2-3}
\cmidrule(l{2mm}r{2mm}){4-5}
\cmidrule(l{2mm}r{2mm}){6-8}
\cmidrule(l{2mm}r{2mm}){9-10}
Channel(s) & CRPS & RMSE & CRPS & RMSE & POD & FAR & TS & CRPS & RMSE \\
\midrule
\multirow{2}{*}{171} & 43.9 & 81.0 & 45.3 & 78.3 & \multirow{2}{*}{0.65} & \multirow{2}{*}{0.18} & \multirow{2}{*}{0.57} & \multirow{2}{*}{44.4} & \multirow{2}{*}{80.1} \\
 & \scriptsize(±1.5) & \scriptsize(±2.7) & \scriptsize(±5.1) & \scriptsize(±9.2) & & & & & \\
\addlinespace
\multirow{2}{*}{193} & 43.5 & 80.0 & 45.5 & 77.7 & \multirow{2}{*}{0.66} & \multirow{2}{*}{0.21} & \multirow{2}{*}{0.56} & \multirow{2}{*}{44.1} & \multirow{2}{*}{79.3} \\
 & \scriptsize(±1.6) & \scriptsize(±2.8) & \scriptsize(±4.7) & \scriptsize(±9.2) & & & & & \\
\addlinespace
\multirow{2}{*}{211} & 41.8 & 77.0 & 47.4 & 80.3 & \multirow{2}{*}{0.66} & \multirow{2}{*}{0.15} & \multirow{2}{*}{\textbf{0.59}} & \multirow{2}{*}{43.5} & \multirow{2}{*}{78.0} \\
 & \scriptsize(±1.5) & \scriptsize(±2.7) & \scriptsize(±4.7) & \scriptsize(±9.2) & & & & & \\
\addlinespace
\multirow{2}{*}{HMI} & 54.6 & 98.3 & 74.3 & 119.7 & \multirow{2}{*}{0.38} & \multirow{2}{*}{0.45} & \multirow{2}{*}{0.29} & \multirow{2}{*}{59.9} & \multirow{2}{*}{104.5} \\
 & \scriptsize(±2.2) & \scriptsize(±3.4) & \scriptsize(±9.3) & \scriptsize(±11.6) & & & & & \\
\addlinespace
\multirow{2}{*}{171-193} & 42.1 & 77.6 & \textbf{42.8} & \textbf{73.5} & \multirow{2}{*}{\textbf{0.67}} & \multirow{2}{*}{0.16} & \multirow{2}{*}{\textbf{0.59}} & \multirow{2}{*}{\textbf{42.3}} & \multirow{2}{*}{\textbf{76.1}} \\
 & \scriptsize(±1.4) & \scriptsize(±2.5) & \scriptsize(±4.3) & \scriptsize(±8.5) & & & & & \\
\addlinespace
\multirow{2}{*}{171-211} & \textbf{41.6} & \textbf{76.9} & 44.9 & 75.7 & \multirow{2}{*}{0.66} & \multirow{2}{*}{0.17} & \multirow{2}{*}{0.58} & \multirow{2}{*}{42.6} & \multirow{2}{*}{76.5} \\
 & \scriptsize(±1.3) & \scriptsize(±2.5) & \scriptsize(±4.4) & \scriptsize(±7.6) & & & & & \\
\addlinespace
\multirow{2}{*}{193-211} & 41.9 & 77.1 & 46.1 & 79.0 & \multirow{2}{*}{0.66} & \multirow{2}{*}{0.15} & \multirow{2}{*}{\textbf{0.59}} & \multirow{2}{*}{43.2} & \multirow{2}{*}{77.7} \\
 & \scriptsize(±1.3) & \scriptsize(±2.6) & \scriptsize(±4.5) & \scriptsize(±9.3) & & & & & \\
\addlinespace
\multirow{2}{*}{171-193-211} & 42.0 & 77.4 & 45.1 & 76.0 & \multirow{2}{*}{0.62} & \multirow{2}{*}{\textbf{0.14}} & \multirow{2}{*}{0.56} & \multirow{2}{*}{43.0} & \multirow{2}{*}{76.9} \\
 & \scriptsize(±1.5) & \scriptsize(±2.7) & \scriptsize(±4.7) & \scriptsize(±8.9) & & & & & \\
\addlinespace
\multirow{2}{*}{171-193-HMI} & 42.5 & 78.4 & 46.1 & 79.1 & \multirow{2}{*}{0.57} & \multirow{2}{*}{0.19} & \multirow{2}{*}{0.50} & \multirow{2}{*}{43.7} & \multirow{2}{*}{78.6} \\
 & \scriptsize(±1.4) & \scriptsize(±2.5) & \scriptsize(±5.2) & \scriptsize(±9.8) & & & & & \\
\addlinespace
\multirow{2}{*}{171-211-HMI} & 41.7 & \textbf{76.9} & 47.5 & 80.3 & \multirow{2}{*}{\textbf{0.67}} & \multirow{2}{*}{0.18} & \multirow{2}{*}{0.58} & \multirow{2}{*}{43.5} & \multirow{2}{*}{78.0} \\
 & \scriptsize(±1.5) & \scriptsize(±2.6) & \scriptsize(±4.7) & \scriptsize(±9.1) & & & & & \\
\addlinespace
\multirow{2}{*}{193-211-HMI} & 41.9 & 77.3 & 44.7 & 77.0 & \multirow{2}{*}{0.65} & \multirow{2}{*}{0.20} & \multirow{2}{*}{0.56} & \multirow{2}{*}{42.8} & \multirow{2}{*}{77.2} \\
 & \scriptsize(±1.4) & \scriptsize(±2.5) & \scriptsize(±4.7) & \scriptsize(±9.4) & & & & & \\
\bottomrule
\end{tabular}
}%
\end{table}

Among the single-channel inputs, the 211 \AA\ channel input leads to the best SWS prediction performance, due to its advantage for the timeline metrics. Although the HSS peak value errors of the 211 model are slightly worse than those of the 171 and 193 models, it achieves the best prediction score metrics of all single-channel models and also yields the best HSS event metrics.
Especially the FAR is smaller, which is likely due to fewer filament channels being misclassified as coronal holes.
The two other AIA channels lead to worse performance in the timeline metrics but facilitate more accurate HSS peak value prediction, although the peak CRPS differences among the single-channel models are small relative to their uncertainty.
We conclude that the 211 \AA\ channel contains the most useful information for the general model accuracy, whereas the 171 \AA\ and 193 \AA\ channels contain better information about HSSs.
The HMI model is a clear outlier. All metrics are substantially worse than those of all other models, well beyond the reported uncertainty, indicating that our model cannot extract useful information from the HMI channel.

This can be physically explained since coronal holes are defined by open coronal magnetic topology, visible directly as intensity deficits in coronal EUV emission (our AIA channels), whereas HMI's photospheric line-of-sight field does not directly express this topology without additional coronal extrapolation (e.g., PFSS modeling), which was likely not learned by the model due to a limited dataset. 
Current HMI data products are further affected by known systematic effects (e.g., limb-distance and magnetic-element-size calibration uncertainties, and the lack of far-side field information), so we do not conclude that magnetograms are generally unhelpful for SWS forecasting, only that they add little value in our specific setup.

When combining the 193 \AA\ with the 211 \AA\ channel, the only effect is a small improvement of the HSS peak value metrics, and the prediction score metrics barely change. 
This suggests that the information in the 193 \AA\ channel overlaps strongly with that of the 211 \AA\ channel. 
In contrast, the 171 \AA\ channel, although performing worse when used as a single input, improves both the timeline and the HSS peak value metrics when combined with one of the two other AIA channels. 
Also, the event detection metrics of the 171-193 model benefit. 
We conclude that the 171 \AA\ channel provides complementary information to the 193 \AA\ and the 211 \AA\ channels. 

Adding a third AIA channel does not further improve the results, besides a small advantage for the addition of the 171 \AA\ channel to the 193-211 model. 
This confirms the information overlap between the 193 \AA\ and 211 \AA\ channels and hints again at the complementary information of the 171 \AA\ channel. 
Also, adding HMI to any two-channel AIA combination does not improve timeline metrics and only yields a slight increase in the HSS peak value metrics for the 193-211 model. 
This again highlights the limited contribution of HMI for SWS forecasting in our setup.

Overall, the combination of 171 \AA\ with one of the two other AIA channels delivers the best performance, with the 171-193 model slightly outperforming the 171-211 model, but we note that differences of the metrics are within the uncertainty bounds. It achieves a timeline CRPS of 42.1~km/s and a timeline RMSE of 77.6~km/s, and a CRPS of 42.8~km/s and RMSE of 73.5~km/s for the HSS peak values, resulting in the best CRPS prediction score of 42.3~km/s and RMSE prediction score of 76.1~km/s. Its POD of 0.67 and FAR of 0.16 yield a TS of 0.59.

Besides the HMI model, differences between channels are generally small, but their ranking is robust against changes in the prediction score weights. To achieve a good performance, it seems sufficient for the model to identify coronal holes in the images, and the specific channel choice mainly fine-tunes the accuracy rather than leading to major improvements. Due to the negative HMI results, we did not test additional combinations of a single AIA channel with HMI.

In summary, we find that the 193 \AA\ and the 211 \AA\ channels contain largely overlapping information, while 171 \AA\ adds complementary features to them. 
The HMI channel does not provide useful information in our approach. Therefore, combinations of two AIA channels yield the best performance, with the 171-193 model outperforming the other combinations.

\subsection{Full Model}
\label{sec:full_model}

The results of the channel study for the full model, including the physical features, are shown in Table~\ref{tab:full_model}. 
Figure~\ref{fig:channel_study}(b) visualizes the corresponding CRPS prediction scores.

\begin{table}[t]
\caption{\label{tab:full_model}Evaluation metrics of full models (using all available inputs, including physical features) for different combinations of solar image and magnetogram inputs. CRPS and RMSE values are in km/s. The 95\% confidence interval half-width is shown in the second row for each model, where available. Bold numbers indicate the best value per column.}
\centering
\adjustbox{max width=\linewidth}{%
\begin{tabular}{lccccccccc}
\toprule
 & \multicolumn{2}{c}{Timeline} & \multicolumn{2}{c}{HSS peak values} & \multicolumn{3}{c}{HSS events} & \multicolumn{2}{c}{Prediction Score} \\
\cmidrule(l{2mm}r{2mm}){2-3}
\cmidrule(l{2mm}r{2mm}){4-5}
\cmidrule(l{2mm}r{2mm}){6-8}
\cmidrule(l{2mm}r{2mm}){9-10}
Channel(s) & CRPS & RMSE & CRPS & RMSE & POD & FAR & TS & CRPS & RMSE \\
\midrule
\multirow{2}{*}{171} & 42.8 & 78.7 & 47.3 & 81.2 & \multirow{2}{*}{\textbf{0.69}} & \multirow{2}{*}{0.16} & \multirow{2}{*}{\textbf{0.61}} & \multirow{2}{*}{44.2} & \multirow{2}{*}{79.5} \\
 & \scriptsize(±1.5) & \scriptsize(±2.7) & \scriptsize(±4.8) & \scriptsize(±9.7) & & & & & \\
\addlinespace
\multirow{2}{*}{193} & 42.9 & 79.1 & \textbf{44.0} & 75.9 & \multirow{2}{*}{0.64} & \multirow{2}{*}{0.18} & \multirow{2}{*}{0.57} & \multirow{2}{*}{43.2} & \multirow{2}{*}{78.0} \\
 & \scriptsize(±1.7) & \scriptsize(±3.1) & \scriptsize(±4.7) & \scriptsize(±10.7) & & & & & \\
\addlinespace
\multirow{2}{*}{211} & \textbf{40.7} & \textbf{75.0} & 48.3 & 83.0 & \multirow{2}{*}{0.66} & \multirow{2}{*}{0.15} & \multirow{2}{*}{0.59} & \multirow{2}{*}{43.0} & \multirow{2}{*}{77.5} \\
 & \scriptsize(±1.4) & \scriptsize(±2.5) & \scriptsize(±5.5) & \scriptsize(±11.2) & & & & & \\
\addlinespace
\multirow{2}{*}{HMI} & 45.7 & 84.2 & 74.7 & 118.2 & \multirow{2}{*}{0.54} & \multirow{2}{*}{0.18} & \multirow{2}{*}{0.48} & \multirow{2}{*}{52.5} & \multirow{2}{*}{93.2} \\
 & \scriptsize(±1.8) & \scriptsize(±3.0) & \scriptsize(±7.0) & \scriptsize(±10.9) & & & & & \\
\addlinespace
\multirow{2}{*}{171-193} & 41.2 & 75.8 & 46.4 & 79.4 & \multirow{2}{*}{0.67} & \multirow{2}{*}{\textbf{0.13}} & \multirow{2}{*}{\textbf{0.61}} & \multirow{2}{*}{42.8} & \multirow{2}{*}{76.9} \\
 & \scriptsize(±1.5) & \scriptsize(±2.8) & \scriptsize(±4.8) & \scriptsize(±8.8) & & & & & \\
\addlinespace
\multirow{2}{*}{171-211} & 41.0 & 75.8 & 45.3 & 77.1 & \multirow{2}{*}{0.68} & \multirow{2}{*}{0.17} & \multirow{2}{*}{0.60} & \multirow{2}{*}{\textbf{42.3}} & \multirow{2}{*}{76.2} \\
 & \scriptsize(±1.3) & \scriptsize(±2.5) & \scriptsize(±4.6) & \scriptsize(±9.3) & & & & & \\
\addlinespace
\multirow{2}{*}{193-211} & 41.2 & 76.0 & 45.0 & 77.8 & \multirow{2}{*}{0.64} & \multirow{2}{*}{0.16} & \multirow{2}{*}{0.57} & \multirow{2}{*}{42.4} & \multirow{2}{*}{76.6} \\
 & \scriptsize(±1.4) & \scriptsize(±2.5) & \scriptsize(±5.6) & \scriptsize(±10.8) & & & & & \\
\addlinespace
\multirow{2}{*}{171-193-211} & 41.4 & 75.9 & 44.7 & \textbf{75.5} & \multirow{2}{*}{0.68} & \multirow{2}{*}{0.17} & \multirow{2}{*}{0.60} & \multirow{2}{*}{42.5} & \multirow{2}{*}{\textbf{75.7}} \\
 & \scriptsize(±1.4) & \scriptsize(±2.7) & \scriptsize(±4.3) & \scriptsize(±9.5) & & & & & \\
\addlinespace
\multirow{2}{*}{171-193-HMI} & 41.6 & 76.8 & 44.8 & 78.1 & \multirow{2}{*}{\textbf{0.69}} & \multirow{2}{*}{0.19} & \multirow{2}{*}{0.60} & \multirow{2}{*}{42.6} & \multirow{2}{*}{77.2} \\
 & \scriptsize(±1.5) & \scriptsize(±2.7) & \scriptsize(±5.0) & \scriptsize(±10.4) & & & & & \\
\addlinespace
\multirow{2}{*}{171-211-HMI} & 41.1 & 76.1 & 46.8 & 80.6 & \multirow{2}{*}{0.68} & \multirow{2}{*}{0.20} & \multirow{2}{*}{0.58} & \multirow{2}{*}{42.9} & \multirow{2}{*}{77.5} \\
 & \scriptsize(±1.4) & \scriptsize(±2.4) & \scriptsize(±5.0) & \scriptsize(±9.8) & & & & & \\
\addlinespace
\multirow{2}{*}{193-211-HMI} & 41.3 & 76.4 & 46.3 & 80.2 & \multirow{2}{*}{0.66} & \multirow{2}{*}{0.16} & \multirow{2}{*}{0.59} & \multirow{2}{*}{42.8} & \multirow{2}{*}{77.7} \\
 & \scriptsize(±1.4) & \scriptsize(±2.6) & \scriptsize(±5.2) & \scriptsize(±10.8) & & & & & \\
\bottomrule
\end{tabular}
}%
\end{table}

All models with physical features improve their timeline metrics as compared to the image-only models. 
However, for the HSS peak value metrics, there is no clear trend: 5 out of 11 models improve their CRPS and 6 out of 11 improve their RMSE. 
In combination, 9 of the 11 models improve both prediction scores. 
For the HSS event detection metrics, we observe a similarly positive trend. The POD increases for most models, accompanied by a slightly higher FAR. In total, 7 out of 11 models improve their TS, 3 remain unchanged, and only one worsens.

Particularly for the HMI model, the timeline, HSS event, and prediction score metrics strongly improve, which indicates that the physical features are highly informative. 
However, since the same features only slightly improve the AIA-based models, it follows that the information content in the physical features strongly overlaps with the information content from the AIA images.
This is plausible since the physical features encode the solar wind from recurrent coronal holes and the state of the solar cycle, which can also partially be inferred from the images. 
Additionally, we conclude that the information provided by the physical features is useful to increase the accuracy of the timeline metrics, but that they do not improve HSS peak predictions. 


The relative ordering of the channels is similar to that of the image-only models (Section~\ref{sec:image-only}). 
Adding a second channel improves the performance of single-channel models, while a third channel or magnetogram supplies redundant information with the other channels and therefore provides no further advantages.
However, including the physical features seems to compensate for some of the differences between the channels, because the model performances are clustered closer together, which can be well seen in Figure~\ref{fig:channel_study}.
The only strong outlier remains the HMI model, although it improves with the physical features.

Notably, the 211 model performs best for the timeline metrics among all models of the study, and the 193 \AA\ model does so for the HSS peak metrics. 
This is consistent with \citeA{Brown22}, who also used the 211 \AA\ channel for their model to optimize the timeline RMSE, and with the ESWF model \cite{Reiss16,Milosic23}, which uses the 193 \AA\ channel for improved HSS peak predictions. 
However, the 193 \AA\ and 211 \AA\ models are not well-balanced between the two metrics, which is why the two-channel models achieve better prediction scores.

The best-performing model again uses a combination of 171 \AA\ with another AIA channel, here being the 171-211 model. 
It has the lowest CRPS prediction score of 42.3~km/s, with a timeline CRPS of 41.0~km/s and timeline RMSE of 75.8~km/s, and an HSS peak value CRPS of 45.3~km/s and RMSE of 77.1~km/s. 
It also has the second-best RMSE prediction score of 76.2~km/s, and its POD of 0.68 and FAR of 0.18 produce a TS of 0.59. 

However, the prediction score differences among the top models (171-211, 193-211, and 171-193-211: 42.3, 42.4, and 42.5~km/s, respectively) are small, and within the uncertainty bounds of timeline and peak metrics. 
We further verified that the relative ranking within this group is sensitive to the specific choice of prediction score weighting. While our default weighting ($w_\text{t}=2, w_\text{p}=1$) favors 171-211, an equal weighting ($w_\text{t}=w_\text{p}=1$) would instead favor 193-211. 
We therefore do not interpret 171-211 as a uniquely superior configuration, but select it as a representative choice among several comparable models, based additionally on its favorable general timeline accuracy.

In summary, we find that incorporating the physical features into the model improves its overall accuracy.
The advantage is consistent, albeit modest, as their informative value overlaps with the content of the input solar images. 
We choose the 171-211 model as a representative choice of the best-performing configurations and use it for all further analysis in this work.
All models and their predictions are publicly available in \citeA{Collin26data}.

\section{Analysis of the Proposed Method}
\label{sec:results}

In this section, we analyze the prediction capabilities of our best model, which uses the 171 \AA\ and 211 \AA\ solar image channels and the available physical features as input.
We show especially the advantage of \sysacronym{} over mean-based models. In Section~\ref{sec:prob_calibration}, we focus on the probabilistic calibration accuracy, in Section~\ref{sec:prob_hss}, we analyze HSS predictions, and in Section~\ref{sec:solar_cycle}, we monitor the performance over the solar cycle. A comparison to other methods is presented separately in Section~\ref{sec:comparison}.

\subsection{Probabilistic Calibration}
\label{sec:prob_calibration}

In this section, we assess the probabilistic calibration of the model, i.e., whether the predicted distributions provide realistic estimates of the uncertainty. 
Although the CRPS already provides a quantitative measure of probabilistic skill, we focus here on a qualitative assessment and the advantages of probabilistic over single-value predictions. 

\begin{figure}[htbp]
\centering
\includegraphics[width=1.0\textwidth]{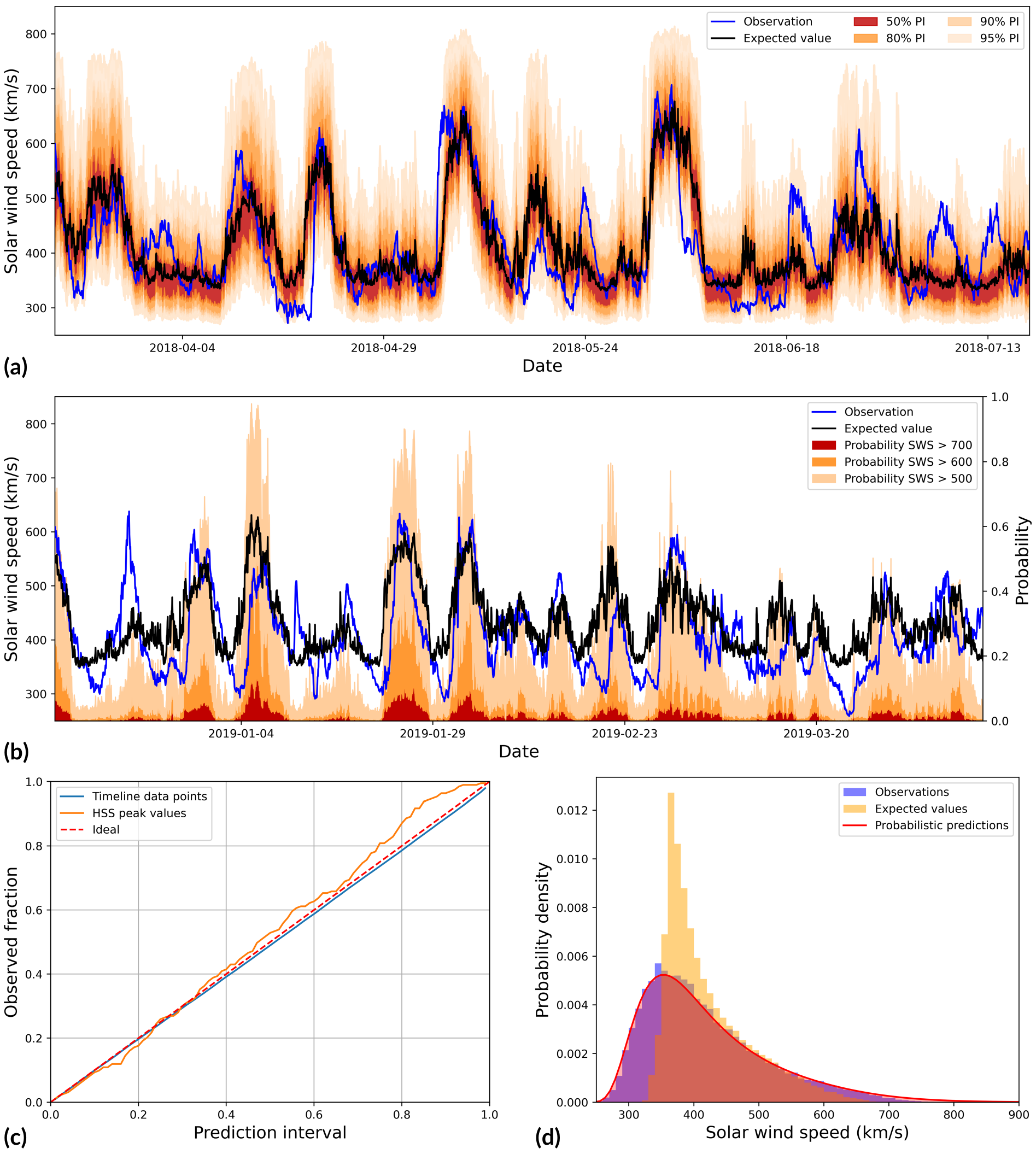}
\caption{
(a) Predicted distributions, shown as expected values with PIs. All deviations of the expected value from the observations are captured within the predicted uncertainty.
(b) Predicted probabilities of the SWS exceeding 500, 600, and 700~km/s, represented by the upper boundary of the corresponding shaded areas. The second y-axis gives the scale of that probability.
These probabilities provide explicit risk quantification beyond single-value predictions. 
(c) Calibration curve for predicted distributions of the timeline and HSS peak values. For PI levels between 0 and 1, the corresponding fraction of observations within each PI is shown. The close agreement with the diagonal indicates very well-calibrated uncertainties.  
(d) Histograms of observations, expected value single-value predictions, and aggregated predicted probability densities. The predicted distributions closely match the observed distribution, while the single-value predictions are biased toward the mean and underestimate the heavy tail of high speeds.}
\label{fig:prob}
\end{figure}

One of the most important aspects of uncertainty quantification is being able to compute prediction intervals (PIs). 
A PI of level $\alpha$ is expected to contain the observation with probability $\alpha$ if the model is well-calibrated.
They can be directly obtained from the quantile function of the predicted distribution, as shown in Section~\ref{sec:prediction}. 
Figure~\ref{fig:prob}(a) shows the predicted expected value and multiple PIs for several HSSs during the declining phase of solar cycle 24. 
The expected value, which is the only output we would get from a mean-based model, captures the time series and the HSS peaks generally well. 
Nevertheless, we also observe deviations from the actual SWS, which are only captured by the additional information that the PIs provide. 
In this example, the 50\% PI overlaps with most smaller deviations, and increasing the interval until the 95\% PI captures nearly all observations, including the peaks of the HSSs which are underestimated by the expected value alone. 

An additional risk quantification is the probability that the SWS exceeds a certain threshold. We obtain this quantity from the predicted CDF, as described in Section~\ref{sec:prediction}.
Figure~\ref{fig:prob}(b) shows an example of probabilities for exceeding 500, 600, and 700~km/s during a time period near solar minimum. 
Besides the one HSS that the model clearly misses, it predicts high probabilities of at least around 80\% for exceeding 500~km/s for the majority of HSSs, medium probabilities around 30\% for exceeding 600~km/s, and low probabilities around 10\% for exceeding 700~km/s. 
The observed HSS peak speeds around 600-650~km/s are slightly higher than one would expect from the predicted probabilites, but generally consistent with the predictions, indicating realistic probabilities. 
The peaks of the expected values are close to the observed ones, but they provide no uncertainty or risk estimates.

We evaluate these predictions using the BSS,
as there is an increasing class imbalance in the dataset for higher thresholds. 
The skill scores are 0.256, 0.154, and -0.095, respectively. This reflects a good skill of the model for the thresholds of 500 and 600 km/s, as it provides a relatively large advantage over naively using the event frequencies in the dataset as probabilities.
However, we also see a decreasing skill for higher SWS values, becoming slightly negative for the threshold of 700 km/s. 
Although this indicates a slight underestimation of the probabilities for the most important data points, these make up a fraction of only 0.6\% in the dataset.
54\% of these points correspond to CMEs, which our model cannot predict, and the other 46\% concentrate around the peak phase of strong HSSs, with a median distance of 11h to the peak.
Thus, statistically, these are outlier points and do not affect the capability of the model to predict the occurrence probabilities of HSSs, as illustrated in the next section.

We can assess the quality of the prediction intervals, and thereby the overall calibration of the predictions, by using Figure~\ref{fig:prob}(c). It shows for all prediction interval levels between 1\% and 99\% the fraction of observations falling into the corresponding PI. 
For a perfectly calibrated model, this fraction of observations is exactly $\alpha$, i.e., the predicted probability of an observation being within the PI. 
For our model, we observe minimal deviations from the ideal diagonal for the timeline predictions. The average absolute deviation is just 0.9\%.
For the subset of predicted HSS peak values, the deviation is slightly larger, but with an average absolute deviation of 2.9\%, it remains small overall. 
These results demonstrate that the uncertainty estimates of our model are highly accurate.

Finally, we assess how well the model predicts the underlying distribution of observations. 
Figure~\ref{fig:prob}(d) compares the histogram of all observations with the histogram of the single-value predictions, i.e., the expected values, and the predicted density of the observations, i.e., the averaged predicted distributions. We see that the single-value predictions are clustered around the mean of the observed distribution and underestimate the heavy tails. This is consistent with findings of previous mean-based models \cite{Upendran20,Bailey21,Raju21,Brown22,Collin25}. In contrast, the probabilistic predictions reproduce the observed distribution almost perfectly.
This shows that the model not only predicts realistic uncertainties for individual time steps but also captures the important data points in the upper tail of the distribution.

In summary, we find that from the probabilistic forecasts, we can extract a variety of useful information for operational SWS prediction. The model predicts realistic probabilities for fast solar wind conditions and very accurately calibrated uncertainties, enabling much more comprehensive forecasts than mean-based (i.e., single-value prediction) methods.

\subsection{High-Speed Streams}
\label{sec:prob_hss}

\begin{figure}[htbp]
\centering
\includegraphics[width=1.0\textwidth]{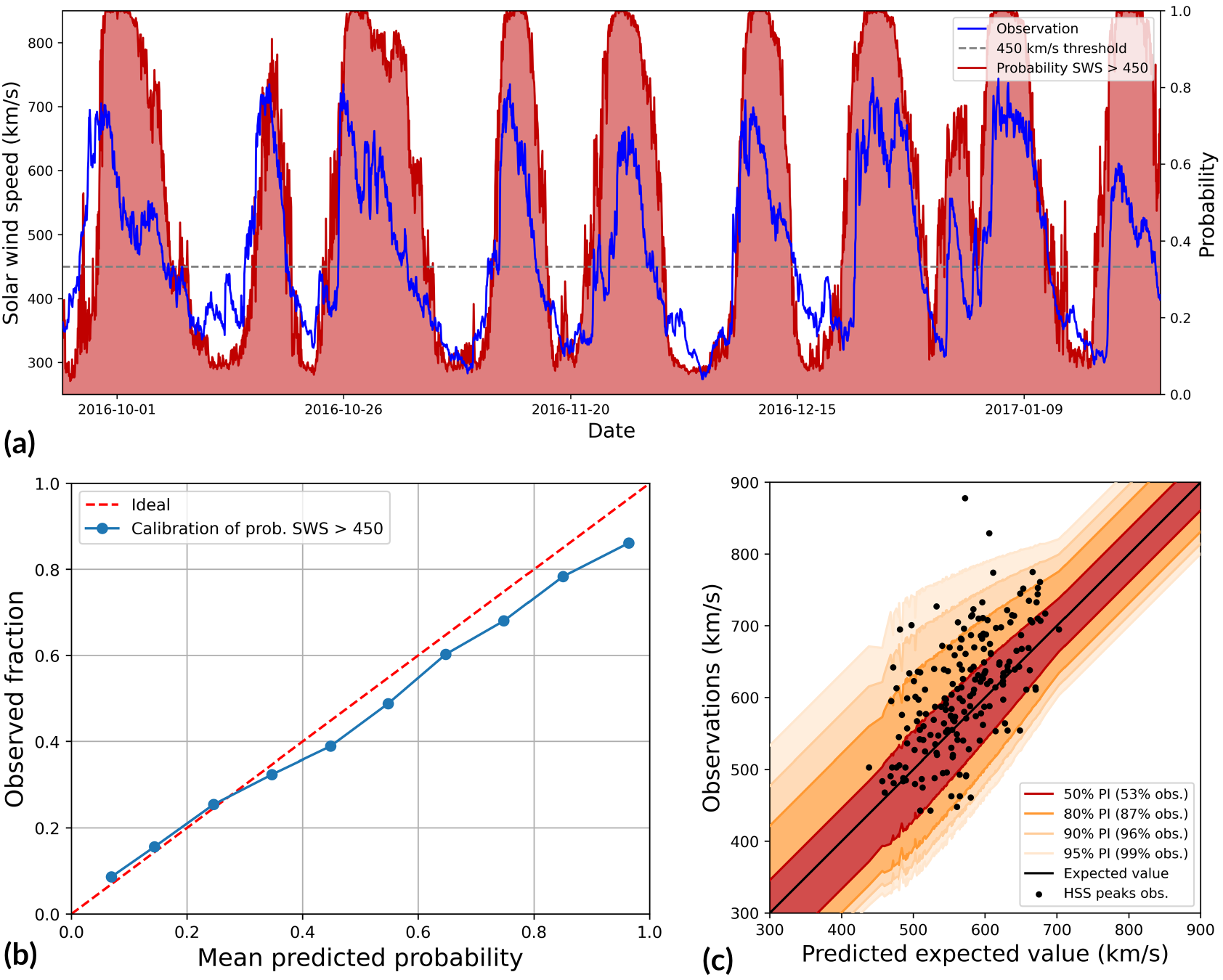}
\caption{(a) Predicted probability that the SWS exceeds 450 km/s, used as a proxy for HSS occurrence. Probabilities above 50\% agree well with observed HSSs and provide a probabilistic criterion for the occurrence.
(b) Calibration curve for the probability that the SWS exceeds 450 km/s. Predictions are grouped into 10\% bins. For each bin, the mean predicted probability and observed event frequency are shown. The close agreement with the diagonal indicates a good calibration for our model.
(c) Predicted vs. observed HSS peak speeds. The predictions are ordered based on the expected value, and the prediction intervals are shown. The actual observations are shown as black dots. The frequency with which observations are contained in a prediction interval is indicated in the legend. The observations scatter around the ideal diagonal line, and, apart from two outliers, the uncertainty is well captured.}
\label{fig:hss}
\end{figure}

In this section, we investigate the capability of the model to forecast the occurrence and peak velocities of HSSs. Similar to the previous section, we derive an HSS occurrence probability from the predicted velocity distributions by computing the probability that the SWS exceeds 450~km/s. Figure~\ref{fig:hss}(a) illustrates this probability for a sequence of HSSs during the declining phase of solar cycle 24. We see that the probabilities closely follow the observed increase of the SWS and remain high until the end of the event, indicating a useful proxy for the occurrence of HSSs. 
The BSS is 0.243, proving a high skill of the model for predicting the occurrence probabilities.

If we classify all time steps with an HSS probability exceeding 50\% as HSSs and compare them with the (ground-truth) HSS list defined in Section~\ref{sec:hss_cme}, we obtain an accuracy of 74\%. 
To further assess the reliability of the predicted probabilities, we compute a calibration curve, shown in Figure~\ref{fig:hss}(b). 
A calibration curve evaluates how well predicted probabilities correspond to the observed event frequencies. 
In contrast to the calibration curve from Figure~\ref{fig:prob}(c), we now evaluate event probabilities instead of PIs.
Thus, to construct the curve, we first group the predicted probabilities into bins of 10\%. 
For each bin, we then compute the average predicted event occurrence probability and the corresponding fraction of observed HSS events. 
Plotting these two quantities against each other yields the calibration curve. 
We notice a good calibration up to $\approx$40\%. 
For larger probabilities, we find a small negative deviation, meaning that the model slightly underestimates the likelihood of HSSs. 
The deviation, however, is at most 10\%, which still indicates a good calibration.

We also evaluate the predicted HSS peak values. 
Figure~\ref{fig:hss}(c) shows, ordered by expected value, all HSS peak predictions and their uncertainties. 
The actual corresponding observations are shown as scattered points.
For perfectly accurate predictions, all observations would be on the diagonal, i.e., equal to the expected values.
The presented results show that the model captures the spread of observed peak values well within the predicted uncertainty bounds. 
We find that the predicted uncertainties are slightly too wide, with 53\% of peaks being observed within the 50\% prediction interval, 87\% within the 80\% prediction interval, 96\% within the 90\% prediction interval, and 99\% within the 95\% prediction interval.
There are two outliers above 800~km/s, which the model does not predict well. 
One corresponds to a CME–HSS interaction where the CME was not labeled in the ICME list, and the other one is an HSS preceded by multiple other HSSs, which likely precondition interplanetary space for the last HSS, leading to this unusually high speed.
We observe a slightly smaller predicted uncertainty for higher expected values, which might be due to the regularization of $\sigma$, limiting the uncertainty from above.
Further, we find that there are more observations above the expected value than below, indicating a slight underestimation of the fastest HSSs.

In summary, we can confirm that our model predicts the probability of the occurrence of HSSs with good accuracy and reliably quantifies the uncertainties of their peak values.

\subsection{Solar Cycle}
\label{sec:solar_cycle}

\begin{figure}[htbp]
\centering
\includegraphics[width=1.0\textwidth]{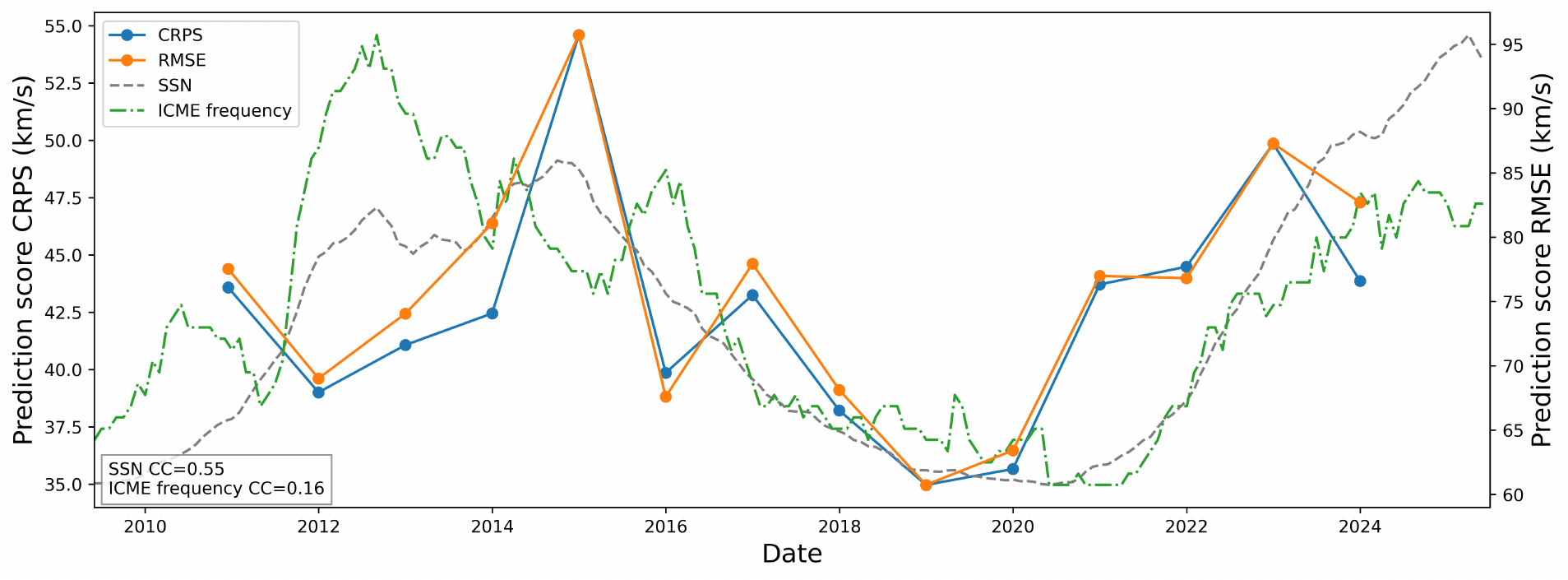}
\caption{Yearly CRPS and RMSE prediction scores through Solar Cycles 24 and 25. 
The smoothed monthly sunspot number (SSN) indicates the phase of the solar cycle. The smoothed monthly ICME frequency indicates the number of CME disturbances of the SWS time series. Both are normalized to the same scale as the prediction scores. The model performance is approximately inversely related to the solar activity, but the correlation with the ICME frequency is low.}
\label{fig:solar_cycle}
\end{figure}

Finally, we analyze the model performance over the solar cycle. 
Figure~\ref{fig:solar_cycle} shows the yearly CRPS and RMSE prediction score values together with the smoothed monthly sunspot number, indicating the phase of the solar cycle. 
We see that both metrics generally follow the trend of the sunspot number, showing a decrease in accuracy with rising solar activity and an increase in accuracy with declining solar activity. 
Accordingly, we find the best model performance during the solar minimum in the years 2018-2020 and the worst performance around the solar maximum of Solar Cycle 24 in 2015. 
Interestingly, the performance remains relatively good during most of the rising phase of Solar Cycle 24, i.e., during the years 2010-2013, and only declines strongly for the peak of the solar activity. 
In contrast, in Solar Cycle 25, the performance starts to decline much earlier, from the start of the rising phase in 2021 onward, consistent with Solar Cycle 25 being a stronger cycle overall, which can be seen by the significantly higher occurrence of sunspots.

To quantify the driver of this solar cycle dependence, we compare the CRPS prediction score against both the smoothed monthly sunspot number and the monthly ICME frequency according to the Richardson and Cane catalog \cite{RCdata}, smoothed using the same 13-month rolling mean as for the sunspot number. We find that the prediction score correlates substantially more  with the sunspot number (CC=0.55) than with ICME frequency (CC=0.16). 
Since ICMEs cause short and high-amplitude prediction errors but do not by themselves explain the broader solar cycle-dependent degradation trend, this indicates that unpredicted CME disturbances alone are not the primary driver of the overall performance decline toward solar maximum. 
Instead, the stronger association with the sunspot number suggests that the global transition of the solar magnetic field from a simple dipolar to a complex multipolar configuration during solar maximum plays a larger role. The increased occurrence of active regions and decreased stability of open field-line coronal holes during this phase likely degrade the predictability of the ambient background solar wind captured by image-based models. We note that this analysis relies on the completeness of the ICME catalog. If the completeness of the catalog itself decreases during high-activity periods, this could partially distort the correlation with the ICME frequency, and we cannot fully exclude a contribution from unidentified CME disturbances to the observed trend.

In summary, we find that our model performs well throughout the solar cycle, but worse at solar maximum than during the declining phase and the solar minimum, driven primarily by the increasing complexity of the solar magnetic field rather than by ICME frequency alone.

\section{Comparison to other Methods}
\label{sec:comparison}

In this section, we compare \sysacronym{} to a selection of simple statistical baseline models, variations of our own model, three machine learning models from the literature, and a physics-based ensemble approach based on HUXt. 
We restrict the comparison to machine learning approaches similar to ours and one physical ensemble model. 
Our analysis should therefore be understood as a comparison of a specific set of approaches, rather than covering all SWS forecasting methodologies. In Section \ref{sec:comparison-main}, we perform our main comparison on data from solar cycle 24, where most models provide predictions. In Section \ref{sec:comparison-wsahuxt}, a second comparison is done to the ensemble approach of HUXt coupled to WSA on a shorter data period of solar cycle 25.

\subsection{Main Comparison on Solar Cycle 24}
\label{sec:comparison-main}

As baseline models, we use simple probabilistic approaches that predict conditional normal distributions with constant standard deviation. 
The climatological model estimates the average of the training data as a constant $\mu$ prediction. 
The persistence model uses the SWS observed 27 days before the target time as the $\mu$ prediction \cite{Bernoux22}. 
In addition, we include a CH baseline model following \citeA{Collin25}, which fits a linear regression model to predict $\mu$, using as input features the SWS measured 27 days earlier and the CH area observed four days earlier in the region between 45° northern and southern latitude around the solar equator and between 30° western and eastern longitude around the central meridian.

To test the core assumptions of our approach, we further evaluate simplified versions of \sysacronym{}. One variant assumes log-normal conditional distributions but uses a constant $\sigma$, removing the variability of the uncertainty. 
Another variant assumes normal conditional distributions with constant $\sigma$, which additionally removes the log-normal assumption.

For comparison to existing work, we include the published predictions from several models in the literature. 
We compare to the gradient boosting model of \citeA{Bailey21}, which uses magnetic model properties as input. 
We further include the Swin Transformer model of \citeA{Brown22}, which is based on 211~\AA\ AIA images and employs a deep learning architecture similar to our $\mu$ prediction pipeline. 
Their predictions are positively biased due to data leakage between their training and test data. 
Nevertheless, we include them because of the similarity of the prediction approaches. 
Additionally, we compare to the polynomial regression model of \citeA{Collin25}, which is based mainly on CH area and SWS persistence features. 
We include both their polynomial output and their second model version that applies a distribution transformation at post-processing to approximate the observed SWS distribution.
Finally, we include predictions from the HUXt model, a reduced-physics approach that produces solar wind flow simulations \cite{Barnard22}. Specifically, we compare to probabilistic hindcasts produced by ensemble simulations from \citeA{Edward-Inatimi24}, where HUXt is coupled with the MAS coronal model, and the probabilistic forecasts produced by ensembles from \citeA{Edward-Inatimi26}, where HUXt is coupled with the WSA coronal model. MAS+HUXt predictions are concatenated simulations of full Carrington rotations, whereas for WSA+HUXt, we concatenate daily updated simulations of the prediction data closest to our four-day lead time. Since the latter provides forecasts only for June 2020 until December 2020 and the full year 2023, which do not overlap with the other literature models, we perform a separate comparison to WSA+HUXt in Section \ref{sec:comparison-wsahuxt}.

Our distributional regression model is, to the best of our knowledge, the first fully probabilistic machine learning model for SWS prediction. 
As the comparison models from the literature are, besides the HUXt models, mean-based, a direct comparison is not straightforward. 
Restricting the comparison to the RMSE of the expected value would neglect the probabilistic information provided by our approach. 
For this reason, we add a comparison in terms of the CRPS by exploiting the implicit assumption of the mean squared error loss, which was used to train all literature models, namely that predictions follow a normal distribution with constant variance. 
Based on this assumption, we extend all single-value prediction models to probabilistic ones by transforming their predictions into conditional normal distributions with mean $\mu$ given by the predicted value and constant $\sigma$. 
The value of $\sigma$ is computed as the standard deviation of the SWS over the 11 years of data preceding the comparison period, corresponding approximately to the length of one solar cycle. 
This ensures that there is no intersection with the test data.

\begin{figure}[htbp]
\centering
\includegraphics[width=0.7\textwidth]{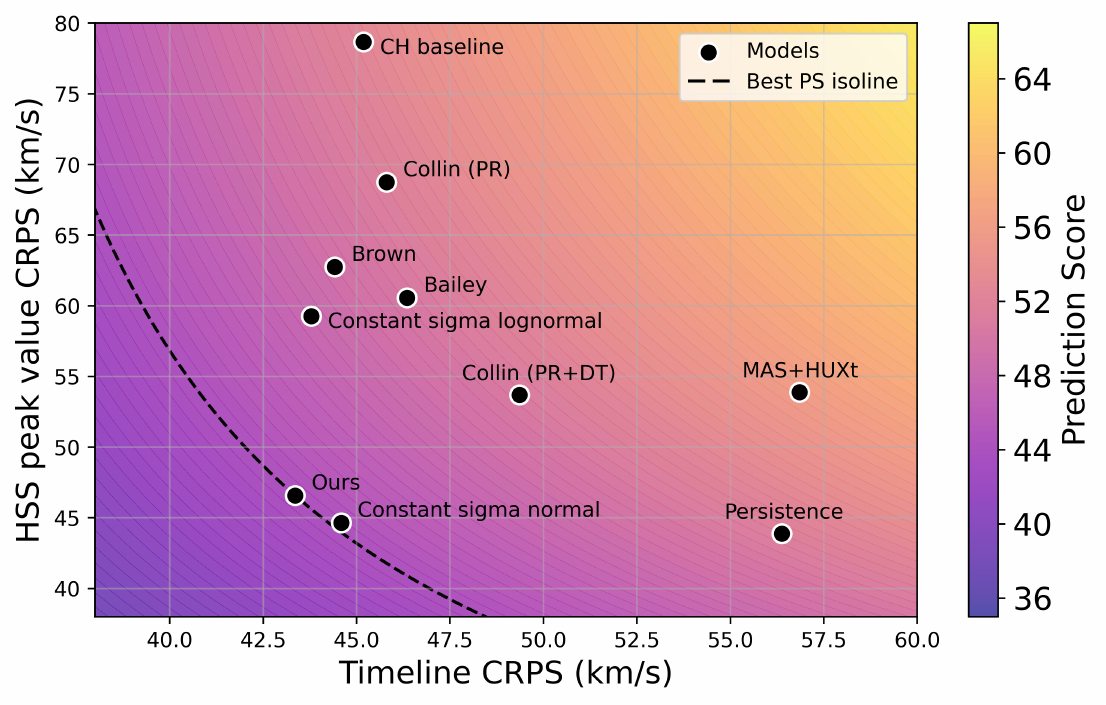}
\caption{Timeline CRPS vs. HSS peak value CRPS of our model compared to other models for the period 2010-2017. 
The trade-off between both errors is visualized by the prediction score, which combines both metrics. 
The dashed curve shows the isoline of the best model's prediction score.
PR = polynomial regression. 
DT = distribution transformation.}
\label{fig:comp}
\end{figure}

We restrict the evaluation to the intersection of available prediction times, ensuring that all models are assessed on the same dataset. 
Small gaps in the predicted time series are linearly interpolated to facilitate HSS filtering.
Most models have few gaps, besides the model of Brown et al., who provide their predictions with frequent gaps of around 4 days.
The resulting evaluation period is June 2010 to October 2017, with some remaining larger gaps of multiple months. The results are summarized in Table \ref{tab:comp} and Figure \ref{fig:comp}.

\begin{table}[th]
\caption{\label{tab:comp}Comparison of our approach to other models, evaluated on the intersection of available dates. 
CRPS and RMSE are given in km/s. The 95\% confidence interval half-width is shown in the second row for each model, where available.
Bold numbers indicate the best values per column. 
PR = polynomial regression. 
DT = distribution transformation.}
\setlength{\tabcolsep}{3pt}
\adjustbox{max width=\linewidth}{%
\begin{tabular}{lccccccccc}
\toprule
 & \multicolumn{2}{c}{Timeline}                  & \multicolumn{2}{c}{HSS peak values}            & \multicolumn{3}{c}{HSS events}                                 & \multicolumn{2}{c}{Prediction Score} \\
\cmidrule(l{2mm}r{2mm}){2-3}
\cmidrule(l{2mm}r{2mm}){4-5}
\cmidrule(l{2mm}r{2mm}){6-8}
\cmidrule(l{2mm}r{2mm}){9-10}
Model (estimator) & CRPS & RMSE & CRPS & RMSE & POD & FAR & TS & CRPS & RMSE \\
\midrule
\multirow{2}{*}{Climatological} & 55.0 & 98.4 & - & - & \multirow{2}{*}{-} & \multirow{2}{*}{-} & \multirow{2}{*}{-} & \multirow{2}{*}{-} & \multirow{2}{*}{-} \\
 & \scriptsize(±3.0) & \scriptsize(±5.4) & & & & & & & \\
\addlinespace
\multirow{2}{*}{CH baseline} & 45.2 & 80.5 & 78.7 & 131.4 & \multirow{2}{*}{0.65} & \multirow{2}{*}{\textbf{0.08}} & \multirow{2}{*}{0.61} & \multirow{2}{*}{52.7} & \multirow{2}{*}{92.4} \\
 & \scriptsize(±1.8) & \scriptsize(±3.6) & \scriptsize(±13.2) & \scriptsize(±16.7) & & & & & \\
\addlinespace
\multirow{2}{*}{Persistence} & 56.4 & 103.0 & \textbf{43.9} & 78.8 & \multirow{2}{*}{0.70} & \multirow{2}{*}{0.26} & \multirow{2}{*}{0.56} & \multirow{2}{*}{51.5} & \multirow{2}{*}{93.4} \\
 & \scriptsize(±2.8) & \scriptsize(±5.1) & \scriptsize(±7.8) & \scriptsize(±14.9) & & & & & \\
\addlinespace
\multirow{2}{*}{Collin PR} & 45.8 & 79.1 & 68.7 & 120.3 & \multirow{2}{*}{0.72} & \multirow{2}{*}{0.12} & \multirow{2}{*}{\textbf{0.66}} & \multirow{2}{*}{51.5} & \multirow{2}{*}{89.3} \\
 & \scriptsize(±1.6) & \scriptsize(±3.8) & \scriptsize(±10.6) & \scriptsize(±16.8) & & & & & \\
\addlinespace
\multirow{2}{*}{Collin PR+DT} & 49.4 & 86.6 & 53.7 & 95.2 & \multirow{2}{*}{\textbf{0.74}} & \multirow{2}{*}{0.14} & \multirow{2}{*}{\textbf{0.66}} & \multirow{2}{*}{50.7} & \multirow{2}{*}{89.3} \\
 & \scriptsize(±2.1) & \scriptsize(±4.1) & \scriptsize(±8.6) & \scriptsize(±17.2) & & & & & \\
\addlinespace
\multirow{2}{*}{Bailey} & 46.3 & 80.2 & 60.6 & 106.9 & \multirow{2}{*}{0.68} & \multirow{2}{*}{0.16} & \multirow{2}{*}{0.60} & \multirow{2}{*}{50.3} & \multirow{2}{*}{87.5} \\
 & \scriptsize(±1.6) & \scriptsize(±3.9) & \scriptsize(±8.9) & \scriptsize(±16.2) & & & & & \\
\addlinespace
\multirow{2}{*}{Brown} & 44.4 & \textbf{76.2} & 62.8 & 110.2 & \multirow{2}{*}{0.69} & \multirow{2}{*}{0.19} & \multirow{2}{*}{0.59} & \multirow{2}{*}{49.2} & \multirow{2}{*}{84.9} \\
 & \scriptsize(±1.5) & \scriptsize(±3.9) & \scriptsize(±9.2) & \scriptsize(±16.4) & & & & & \\
\addlinespace
\multirow{2}{*}{MAS+HUXt} & 56.9 & 97.4 & 53.9 & 97.3 & \multirow{2}{*}{0.69} & \multirow{2}{*}{0.22} & \multirow{2}{*}{0.58} & \multirow{2}{*}{55.8} & \multirow{2}{*}{97.3} \\
 & \scriptsize(±4.1) & \scriptsize(±6.0) & \scriptsize(±9.7) & \scriptsize(±17.5) & & & & & \\
\addlinespace
\multirow{2}{*}{Ours w/ const. $\sigma$, log-normal} & 43.8 & 79.7 & 59.3 & \textbf{78.7} & \multirow{2}{*}{0.72} & \multirow{2}{*}{0.22} & \multirow{2}{*}{0.60} & \multirow{2}{*}{48.0} & \multirow{2}{*}{\textbf{79.4}} \\
 & \scriptsize(±2.2) & \scriptsize(±3.8) & \scriptsize(±5.9) & \scriptsize(±12.2) & & & & & \\
\addlinespace
\multirow{2}{*}{Ours w/ const. $\sigma$, normal} & 44.6 & 79.7 & 44.6 & \textbf{78.7} & \multirow{2}{*}{0.72} & \multirow{2}{*}{0.22} & \multirow{2}{*}{0.60} & \multirow{2}{*}{44.6} & \multirow{2}{*}{\textbf{79.4}} \\
 & \scriptsize(±1.9) & \scriptsize(±3.8) & \scriptsize(±6.7) & \scriptsize(±12.2) & & & & & \\
\addlinespace
\multirow{2}{*}{Ours} & \textbf{43.4} & 79.7 & 46.6 & \textbf{78.7} & \multirow{2}{*}{0.72} & \multirow{2}{*}{0.22} & \multirow{2}{*}{0.60} & \multirow{2}{*}{\textbf{44.4}} & \multirow{2}{*}{\textbf{79.4}} \\
 & \scriptsize(±2.2) & \scriptsize(±3.8) & \scriptsize(±7.2) & \scriptsize(±12.2) & & & & & \\
\bottomrule
\end{tabular}
}%
\end{table}

For timeline predictions, our full model outperforms all other models, achieving a CRPS of 43.4~km/s. 
The improvement over the climatological and persistence baselines is particularly large, i.e., around 10 km/s for the CRPS and 20 km/s for the RMSE. 
We also outperform the Swin Transformer of \citeA{Brown22}. 
Their model, however, has a better RMSE (only 3.5 km/s difference), meaning that the locations of their predicted probability distributions are more accurate. This naturally also leads to an advantage in the CRPS for their model. 
The fact that our CRPS is still smaller indicates higher probabilistic accuracy, i.e., sharper predicted distributions, despite the slightly worse RMSE.
MAS+HUXt performs considerably worse than all machine learning models, including the baselines, with a timeline CRPS of 56.9~km/s, comparable to the persistence baseline.
The model with the second-best CRPS is the log-normal model with constant $\sigma$, which outperforms the normal model with constant $\sigma$, confirming that the log-normal assumption is beneficial for probabilistic SWS modeling. 
The CH baseline model performs surprisingly well (45.2 km/s CRPS and 80.5 km/s RMSE), given its simplicity, and is inferior only to the Swin Transformer models.
We note that, given the 95\% confidence intervals of the errors reported in Table~\ref{tab:comp}, the differences between the top models with respect to the timeline errors are around the order of the error uncertainty, meaning that these results should be interpreted with caution rather than as absolute rankings.

For HSS peak value predictions, the persistence model performs best, with a CRPS of 43.9~km/s. 
The model benefits from not having a bias to underestimate the HSS peak values. 
The second-best model is the normal model with constant $\sigma$ (44.6 km/s), which slightly outperforms our full model (46.6 km/s), although both achieve the same HSS peak RMSE. 
This suggests that HSS peak values are better represented by normal than by log-normal distributions. 
The substantially worse performance of the log-normal model with constant $\sigma$ confirms that the log-normal assumption breaks down for HSS peaks, due to the unphysically long tail of the log-normal distributions. 
Our full model mitigates this issue through regularization of the predicted $\sigma$, restricting it to physically realistic values. 
Even when ignoring the probabilistic aspect, \sysacronym{} strongly improves the prediction of the expected value compared to previous models from the literature, achieving an HSS peak RMSE of 78.7~km/s.
MAS+HUXt performs relatively well, with an HSS peak CRPS of 53.9~km/s, outperforming the CH baseline and other machine learning models besides our own model variants.

For the HSS event detection metrics, differences between models are small. 
The distribution transformation model of \citeA{Collin25} achieves the best performance with a TS of 0.66, but all models perform comparably well in this category.

For the prediction score metrics, there is a clear trend. 
The models from the literature outperform the baseline models, with the exception of MAS+HUXt, while all variants of \sysacronym{} outperform the literature models.
MAS+HUXt achieves the worst prediction score of all compared models, reflecting its comparatively poor timeline performance despite good HSS peak accuracy.
Our full model achieves the best overall performance, with a CRPS prediction score of 44.4~km/s and an RMSE prediction score of 79.4~km/s. 
This demonstrates that our approach balances the competing prediction objectives well and that the combination of log-normal conditional distributions with variable uncertainty provides a benefit for probabilistic SWS forecasts, with a substantial improvement over existing models. 
The variants of our model with constant $\sigma$ also outperform all baseline models, showing that the advantage of our approach is not only due to variable uncertainty, but also due to more accurate mean predictions. 
Notably, the prediction score advantage of the full model over the normal model with constant $\sigma$ is relatively small. 

The comparison across the three variants of our model also isolates the specific contribution of the distributional regression framework, independent of the model architecture, the choice of physical features and image channels, which are held fixed across all three and whose individual contributions are assessed separately (model architecture in \citeA{Brown22}, physical features in \citeA{Collin25}, image channels in Section~\ref{sec:image_study}). 
Moving from a constant $\sigma$ log-normal model to our full model with additionally input-dependent $\sigma$ leaves the timeline CRPS nearly unchanged (43.8 to 43.4~km/s) but substantially improves the HSS peak CRPS (59.3 to 46.6~km/s, a reduction of 21\%) and the prediction score CRPS (48.0 to 44.4~km/s). 
This indicates that the primary benefit of explicitly modeling variable uncertainty is not improved average accuracy, which is instead driven mainly by the underlying architecture and input features, but substantially improved calibration and peak-value performance, which is consistent with the goal of probabilistic forecasting that motivates this work.

In summary, \sysacronym{} outperforms all models in our comparison, providing the most accurate probabilistic SWS forecasts.
Within the group of literature models, our model is the only one without a negative trade-off between timeline and HSS peak performance, thereby achieving a more actionable forecast than the comparison models.
The results support our main assumption that the SWS is better modeled by log-normal conditional distributions with variable uncertainty than by normal distributions with constant uncertainty. 
At the same time, we find that HSS peak values are more accurately described by normal distributions, suggesting to model peaks separately.

\subsection{Comparison to HUXt on Solar Cycle 25}
\label{sec:comparison-wsahuxt}

Finally, we compare our full model to WSA+HUXt, over the shorter period of 1.5 years in 2020 and 2023 in which WSA+HUXt predictions are available. Although the comparison period is short and can therefore not be viewed as fully representative, this might be the most interesting comparison, as WSA+HUXt yields actual probabilistic forecasts obtained from a physical simulation model that is routinely used for operational forecasting.
We find that our model achieves a better timeline accuracy (CRPS of 40.9 (±3.2) vs.~60.8~km/s (±5.5)) and substantially better calibrated prediction intervals (mean absolute deviation of 0.02 vs.~0.28) than WSA+HUXt, while WSA+HUXt achieves better HSS peak accuracy (CRPS of 43.4 (±16.6) vs.~76.0~km/s (±19.6)). 
For the HSS event metrics, both models have comparable PODs (ours of 0.58 vs.~WSA+HUXt of 0.54), while WSA+HUXt has an exceptionally high FAR of 0.48 compared to our FAR of 0.07, resulting in a TS of 0.36 for WSA+HUXt and a TS of 0.56 for our model.
The overall CRPS prediction score favors our model in this comparison (prediction score of 48.4 vs.~53.7~km/s).
These results highlight two fundamentally different weaknesses of both models. While our model underestimates most HSS during that time, resulting in the high peak error, WSA+HUXt contrarily overpredicts the frequency of HSSs, resulting in the high timeline error and FAR while having a high accuracy for peak value predictions when it correctly predicts a HSS occurrence.
However, we emphasize not to over-interpret these results, given the large uncertainties of the peak metrics (±16.6 and ±19.6~km/s) for both models in this short evaluation window.

We summarize the direct comparison between the two HUXt models and our model with the finding that the predicted physical solar wind trajectories of HUXt enable accurate HSS peak predictions, suffering less from the underestimation bias of our model, but with a tendency to overestimate the frequency of HSSs. On the other hand, our probabilistic predictions are on average substantially more accurate and the predicted uncertainties significantly more reliable.

\section{Application of PROSWIN to New Unseen Data}
\label{sec:unseen}

\begin{figure}[htbp]
\centering
\includegraphics[width=1.0\textwidth]{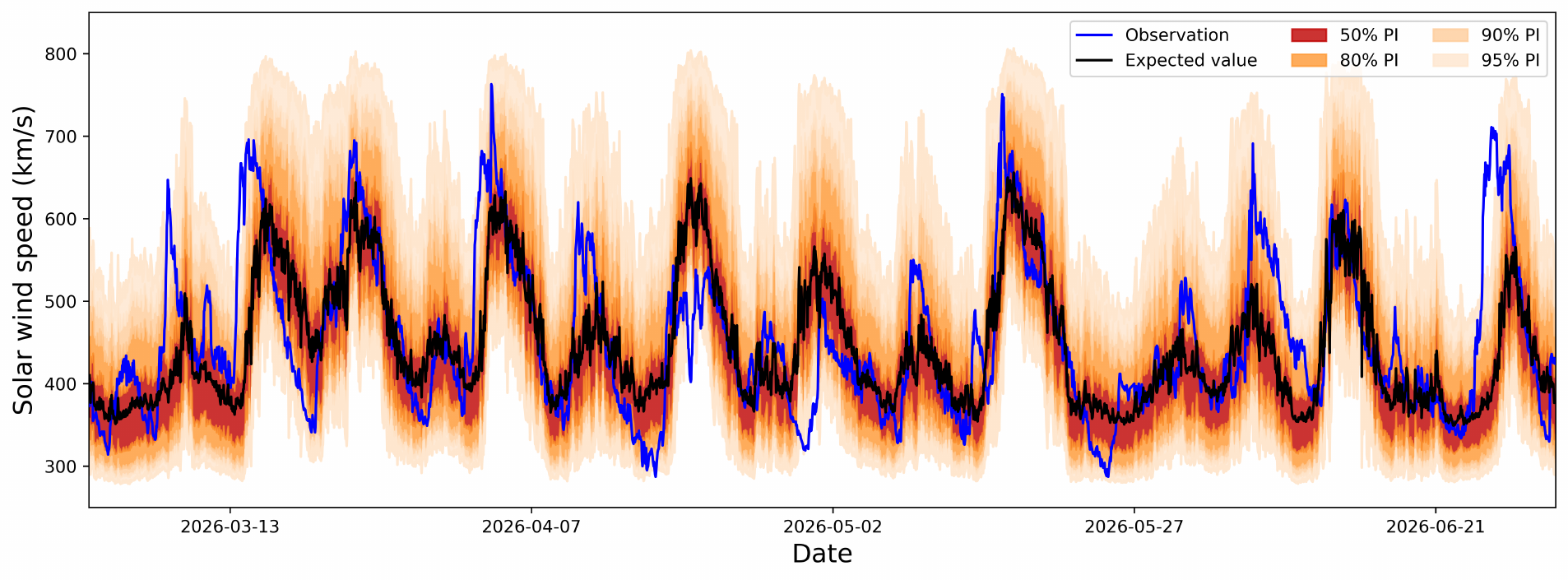}
\caption{Probabilistic predictions of one of the five cross-validation models on new data. The observations are predicted well, indicating a good generalization of the model to new data.}
\label{fig:unseen}
\end{figure}

As a final experiment, we test whether the models which we have developed and cross-validated on data from June 2010 until June 2024 transfer to new unseen data. For that purpose, we use the dataset extension of two years from July 2024 until June 2026, which was excluded from all model development, training, hyperparameter optimization, and previous comparisons. This dataset encompasses 17,425 data points and we use it as input to produce predictions with the five models trained in the cross-validation run of the 171-211 model analyzed throughout this work. The models are not retrained or fine-tuned. An example of the obtained predictions can be seen in Figure \ref{fig:unseen} and the evaluation results are shared in Table \ref{tab:unseen}, compared to the climatological and 27-day persistence baselines.

A qualitative assessment of the predicted time series in Figure \ref{fig:unseen} yields a good performance, as the model predictions capture the majority of SWS enhancements very well, and even underestimated HSS peaks are well within the predicted uncertainty intervals. Quantitatively, all five models outperform both baselines on the timeline metrics and on the combined prediction score, with the weakest fold still achieving a timeline CRPS of 50.5 km/s and a CRPS prediction score of 59.2~km/s, compared to 59.8 km/s and 62.1~km/s, respectively, for persistence.
The predicted uncertainties remain well calibrated, with a mean absolute deviation of the prediction intervals of 2.4\% (range 1.8\%-3.3\% across folds), just slightly worse than the 0.9\% we reported on the dataset until 2024, and essentially unchanged from the value of 2.4\% obtained for the benchmark comparison period until 2017 (Section~\ref{sec:comparison}).
Consistent with our previous findings, the persistence baseline again achieves the best HSS peak accuracy, and our model maintains its underestimation bias for HSS peak values, which is more pronounced in this period, with the best model giving an HSS peak CRPS of just 82.2 km/s.
The Brier skill scores are positive across all evaluated thresholds, ranging from 0.28 for the threshold of 500~km/s, 0.17 for 600~km/s, to 0.07 for 700~km/s. Overall, these results confirm a high prediction skill of the probabilistic forecasts on unseen data, besides the known underperformance against the persistence model for the HSS peaks.

\begin{table}[th]
\caption{\label{tab:unseen}Performance on the unseen data from July 2024 to June 2026. 
For our model, the second row shows the range across the five cross-validation fold models.
CRPS and RMSE are given in km/s. 
Bold numbers indicate the best value per column.}
\centering
\setlength{\tabcolsep}{3pt}
\adjustbox{max width=\linewidth}{%
\begin{tabular}{lccccccccc}
\toprule
 & \multicolumn{2}{c}{Timeline} & \multicolumn{2}{c}{HSS peak values} & \multicolumn{3}{c}{HSS events} & \multicolumn{2}{c}{Prediction Score} \\
\cmidrule(l{2mm}r{2mm}){2-3}
\cmidrule(l{2mm}r{2mm}){4-5}
\cmidrule(l{2mm}r{2mm}){6-8}
\cmidrule(l{2mm}r{2mm}){9-10}
Model & CRPS & RMSE & CRPS & RMSE & POD & FAR & TS & CRPS & RMSE \\
\midrule
\multirow[t]{2}{*}{Climatological} & 62.4 & 115.8 & - & - & \multirow[t]{2}{*}{-} & \multirow[t]{2}{*}{-} & \multirow[t]{2}{*}{-} & \multirow[t]{2}{*}{-} & \multirow[t]{2}{*}{-} \\
 & & & & & & & & & \\
\addlinespace
\multirow[t]{2}{*}{Persistence} & 59.8 & 109.9 & \textbf{67.3} & \textbf{120.0} & \multirow[t]{2}{*}{\textbf{0.76}} & \multirow[t]{2}{*}{\textbf{0.17}} & \multirow[t]{2}{*}{\textbf{0.66}} & \multirow[t]{2}{*}{62.1} & \multirow[t]{2}{*}{113.1} \\
 & & & & & & & & & \\
\addlinespace
\multirow[t]{2}{*}{Ours} & \textbf{47.8} & \textbf{89.0} & 82.2 & 129.8 & 0.68 & 0.22 & 0.57 & \textbf{55.5} & \textbf{99.4} \\
 & \scriptsize[46.5--50.5] & \scriptsize[87.2--92.6] & \scriptsize[69.7--91.1] & \scriptsize[117.6--140.3] & \scriptsize[0.60--0.72] & \scriptsize[0.18--0.29] & \scriptsize[0.48--0.62] & \scriptsize[52.7--59.2] & \scriptsize[96.5--104.3] \\
\bottomrule
\end{tabular}
}%
\end{table}

The absolute errors in this period are larger than the values reported for our main dataset.
This is expected, as the evaluation period falls entirely within the maximum phase of solar cycle~25, the regime in which our model performs worst. Even more, given the relationship between the sunspot number and the model performance found in Section \ref{sec:solar_cycle} and the fact that solar cycle 25 is substantially stronger than solar cycle 24, this explains the decrease in performance.
For comparison, during the maximum of solar cycle~24 (July 2014 to June 2015), our cross-validated model achieves a CRPS prediction score of 54.6~km/s and an RMSE prediction score of 95.7~km/s, compared to 55.5~km/s and 99.4~km/s on the unseen data.
The performance in the two years from the maximum of solar cycle 25 is therefore comparable to, and only marginally worse than, the cross-validated performance during an equivalent but weaker solar maximum of the previous solar cycle.
While this does not rule out a selection bias of the models from the hyperparameter optimization, it indicates that any such bias is small and that the generalization capability of our model is not compromised and estimated realistically by our cross-validation results.

\section{Discussion and Limitations}
\label{sec:discussion}

The ultimate goal of any prediction framework should be operational application to provide a basis for real-world decisions by stakeholders.
We demonstrate how probabilistic models provide important information to support this decision process, such as prediction intervals and probabilities for hazardous events.
Mean-based single-value prediction models do not provide such information, making them less applicable in operational settings. 
While \sysacronym{} is not intended as a standalone forecast for decision-making, its probabilistic SWS predictions could serve both as a general indicator of upcoming space weather risk and as input to downstream operational and research models that depend on hourly SWS averages.
Examples of downstream models include real-time plasmasphere density models (e.g., PINE-RT; \citeNP{Bianco23}), operational Kp index forecasting models (e.g., \citeNP{Shprits19}, \citeNP{Zhelavskaya19}), CME arrival time prediction models (e.g., \citeNP{Liu18}, \citeNP{Yang23}, \citeNP{Chierichini24}, \citeNP{Li24}), and radiation belt models (e.g., ORIENT-R; \citeNP{Chu21}).
These models could significantly profit from having access to reliable SWS distributions with a lead time of four days, as this enables Monte Carlo sampling from the predicted distribution to generate ensembles of the downstream target application, and represents one necessary component towards extending their effective forecast lead time.
However, probabilistic predictions are only useful if they are both accurate and well calibrated, which can be competing objectives \cite{Camporeale21accrue,Camporeale25flare}. 
Developing models that balance accuracy and calibration can be challenging, and it is an important direction to further improve space weather forecasts.

We further note that the hourly cadence of our forecasts inherently limits their applicability to use cases that themselves operate on hourly or coarser timescales, such as the examples listed above. Applications requiring sub-hourly resolution are outside the scope of the present work.

Besides its usage in downstream applications, reliable SWS uncertainty is also relevant for the determination of the geoeffectiveness of space weather events. Alongside the interplanetary magnetic field (IMF) orientation (dominated by the southward component $B_z$), the SWS is one of the two primary drivers of solar wind-magnetosphere coupling \cite{Newell07}. 
This means that a SWS forecast alone is incomplete for geomagnetic disturbance prediction, because a fast HSS with predominantly northward $B_z$ may produce only modest geomagnetic activity, whereas a slower stream with sustained southward $B_z$ can be considerably more geoeffective. 
However, reliable forecasting of $B_z$ beyond very short lead times remains an open and largely unsolved challenge \cite{Reiss21}. 
Given this asymmetry, a well-calibrated SWS forecast still provides meaningful operational value on its own. Since geoeffectiveness scales jointly with speed and IMF orientation, low forecast SWS provides a reliable indicator of low geomagnetic risk, while high forecast SWS indicates a necessary, though not sufficient, precondition for strong geoeffective coupling. 
Therefore, an accurate probabilistic SWS forecast is an important first step, complementary to future efforts addressing $B_z$ prediction.

Within the described scope, we see the specific contribution of our work as follows. 
Unlike physics-based numerical models, our approach does not rely on simplifying physical assumptions, avoiding systematic biases associated with such approximations. 
Our approach further maps directly from solar imagery to Earth-arrival SWS distributions, without requiring a separately derived inner boundary condition, itself a source of error in propagation-based models. We therefore learn the uncertainty directly from the data itself instead of relying on perturbations of boundary conditions leading to correct uncertainties.
Our direct comparison to HUXt (Section~\ref{sec:comparison}) provides quantitative support for these advantages, showing improved accuracy and substantially better calibration. 
We therefore position \sysacronym{} not as a general improvement to solar wind forecasting, but specifically as a computationally efficient, well-calibrated probabilistic alternative for applications operating at hourly to multi-day timescales.

In this work, we demonstrate that utilizing log-normal uncertainties is an effective way to improve probabilistic SWS forecasts, as they are well-suited to model the skewed and heavy-tailed distribution of the ambient SWS. 
This improvement, however, does not transfer to HSS peak velocities, whose uncertainty is better represented by normal distributions. 
This finding suggests the use of mixture models, combining one distribution for quiet-to-moderate solar wind conditions and another for upper-tail events. 
This approach could also help to reduce the underestimation bias of rare hazardous events and would remove the need to clip the predicted variance of log-normal distributions to avoid unphysically high uncertainty intervals.

A fundamental difficulty for most machine learning models in space weather is the strong class imbalance of datasets. 
While quiet times and moderately disturbed periods appear with high frequency, fast solar wind conditions, which are most important for space weather hazards, occur relatively rarely. 
Standard training algorithms minimize the overall error of the entire dataset and therefore lead to models that perform well for average conditions, but neglect rare conditions in the upper tail of the distribution, systematically underestimating important events. 
Although we increase the training weights for HSSs to mitigate this issue, a small bias remains in our model. However, we can still capture strong events well by using distributional regression, which models rare events through higher quantiles of the predicted distribution. 
Statistically, these events are outliers in the dataset and are thereby predicted with lower occurrence probabilities. 
While this does not resolve the underlying issue, it represents one way to address it. Ultimately, however, the goal should be to entirely remove the underestimation bias of fast solar wind conditions, which requires further methodological developments for imbalanced regression scenarios \cite{Camporeale25paris}.

Closely related to this issue is the trade-off between timeline accuracy and the HSS peak accuracy.
Forcing a model to predict rare conditions more accurately typically implies stronger deviations from the mean across the entire dataset, which increases the timeline error. 
This leads to competing optimization objectives, which we observe in our experiments and which was also reported in previous studies (e.g., \citeNP{Shprits19,Collin25}). 
Consequently, it is not sufficient to compute errors averaged over the full dataset alone, as it is typically done by solar wind prediction models. 
To address this, we introduce a new metric that balances these competing objectives by combining timeline and peak errors using a weighted harmonic mean, which we refer to as prediction score.
Operationally, the prediction score should be interpreted as a model selection criterion rather than as a directly actionable forecast error in the way CRPS or RMSE are.
The selected models show good performance across all solar wind conditions, rather than models that optimize one objective at the expense of the other, and we can confirm that our model is significantly better balanced in comparison to previously published SWS models.
However, as we developed it for the assessment of HSS predictions, the metric relies on the previous association of predicted and observed HSSs. 
Further, it does not take into account the detection rate. 
The metric may be extended to address these weaknesses and to generalize to settings where explicit event detection and association are not required, for example, by using a conditional RMSE or CRPS, as suggested by \citeA{Camporeale25paris}.

A constraint of our work is that we trained our model with definitive level 1 solar images rather than near-real-time data.
Although all needed input data is either available in real-time or has an equivalent near-real-time replacement, meaning that an operational implementation would be possible, this means that our work is more a test of potential than an actual assessment of the operational forecast quality.

Relatedly, our hyperparameter optimization procedure, evaluated on the same cross-validation splits used for our reported performance, carries an inherent risk of selection bias \cite{Cawley10}. 
While computational constraints prevented a fully nested cross-validation scheme, our evaluation on a new unseen data period (Section~\ref{sec:unseen}) showed a model performance consistent with expectations given the solar cycle phase, providing evidence that this bias, if present, is small.

A further consequence of the architecture of \sysacronym{} is that it predicts marginal distributions independently at each time point, rather than a jointly consistent forecast trajectory, unlike physical propagation models (e.g., HUXt; \citeNP{Barnard22}) or machine learning models that model explicit temporal dependencies (e.g., Gaussian Process approaches; \citeNP{Rasmussen04}).
We find that prediction residuals are substantially autocorrelated across consecutive hours (the autocorrelation for lags of 1, 12, and 24 hours is 0.97, 0.76, and 0.56, respectively), indicating that independent per-timestep sampling would discard existing temporal structure in the model errors. 
This suggests that a Gaussian-copula-based trajectory sampling approach, coupling the marginals according to this measured dependence structure \cite{Pinson12,Moeller13}, could offer a practical route to generating temporally coherent sample trajectories without modifying the predicted marginal distributions or any reported metric. 
Such a strategy would, however, capture only statistical, pairwise dependence, without guaranteeing that sampled trajectories correspond to physically realistic scenarios. 
Extensions of our data-driven approach to employ explicit time series models or to complement physical simulation models are natural directions for future work.

Finally, \sysacronym{} does not explicitly consider CMEs, as no direct information about CME occurrence or properties is provided as input. 
This restricts the operational applicability, as CMEs are responsible for the most severe geomagnetic storms. 
We note, however, that this does not necessarily mean CME-related information is entirely unavailable to the model. 
Active regions, a weak proxy for CME source likelihood, are clearly visible in the AIA channels and HMI magnetograms used as input. As a black-box model, we cannot rule out that some of that signal is implicitly exploited. 
We tested this by comparing the predicted $\sigma$ around ICME arrivals against non-ICME periods. 
We find a small increase in $\sigma$ of around 3\% during ICME disturbance times in comparison to non-ICME periods, providing weak support for the interpretation that  CME effects manifest as elevated uncertainty.
This is consistent with our channel study (Section~\ref{sec:image_study}), where including HMI magnetograms, a higher-fidelity proxy for CME-relevant magnetic complexity, did not improve performance.

Incorporating CME predictions into machine learning SWS predictions remains a major challenge. 
CMEs are primarily detected in coronagraph data, where parameter estimates are highly uncertain. 
As a result, reliable machine learning CME forecasting tools are rare and have been evaluated only on relatively small datasets (e.g., \citeNP{Liu18,Yang23,Chierichini24,Li24}).
This highlights a structural advantage of physics-based propagation models, such as HUXt, in the context of CME forecasting. CMEs can be incorporated comparatively easily as simple parametrized injections.
Developing a unified data-driven SWS forecast that combines HSSs and CMEs, therefore, is much harder and remains an open problem for future research.

\section{Conclusion}
\label{sec:conclusion}

In this study, we show that solar wind speed forecasting can be done in a fully probabilistic way.
We developed \sysacronym{}, a probabilistic machine learning approach to forecast calibrated SWS distributions. 
Our approach combines a Swin Transformer, which extracts features from multiple SDO solar image channels, with physical input features encoding the solar wind conditions from the previous solar rotation and the current state of the solar cycle.
These features are processed by a neural network coupled to a distributional regression framework, which predicts the distributional parameters $\mu$ and $\sigma$ of a log-normal conditional distribution of the SWS four days in advance. 
This allows a quantification of forecast uncertainty and the risk of fast solar wind conditions, which is of high relevance for operational solar wind forecasting and represents a substantial advantage compared to previous mean-based models.

To analyze \sysacronym{}, we defined a new class of error metrics called prediction score, which takes into account both timeline and HSS peak performance, and is designed to overcome the limitations of standard error metrics, which largely neglect rare but important conditions.
Evaluation of our approach led to the following main results:
\begin{enumerate}
    \item Our representatively selected best model uses 171~\AA\ and 211~\AA\ solar image input, although error differences to other models are small. 
    The 171-211~\AA\ model achieves a CRPS prediction score of 42.3~km/s for the years 2010 to 2024, based on a timeline CRPS of 41.0~km/s and an HSS peak value CRPS of 45.3~km/s. 
    The expected value of the predicted distributions has a timeline RMSE of 75.8~km/s and an HSS peak value RMSE of 77.1~km/s.
    \item \sysacronym{} outperforms the selected benchmark models for the combined prediction score metric and is the only model that does not have a negative trade-off between timeline and HSS peak performance.
    We find that optimizing for the prediction score leads to a significant improvement of the HSS peak metrics and results in more applicable solar wind prediction models. Interestingly, the model consistently achieving the best HSS peak value performance is a simple 27-day persistence baseline. Additionally, WSA+HUXt achieves a similar HSS peak performance on a shorter, non-overlapping period of solar cycle 25.  
    \item The predicted uncertainties are very well calibrated. Prediction intervals deviate on average by only 0.9\% from the observed frequencies, and the aggregated predicted distribution closely matches the observed SWS distribution.
    \item The risks of HSSs are accurately quantified. 
    The model predicts HSS occurrence probabilities with errors of at most 10\% compared to the observed frequencies. 
    For the peak values, the predicted prediction intervals deviate by only 2.9\% on average from the observed frequencies.
    \item Our results indicate that a large overlap of the information which is important for SWS prediction is contained in the 193~\AA\ and 211~\AA\ solar image channels, whereas the 171~\AA\ channel seems to add complementary information to the previous ones. 
    Adding physical features consistently improves model performance, albeit only by a small margin, indicating that a large fraction of the information contained in the used physical features is already encoded in the images.
    \item The SWS is better described by log-normal conditional distributions with variable uncertainty than by normal distributions with constant uncertainty. 
    At the same time, HSS peak values are more accurately represented by normal distributions.
    \item The model performs best during the declining phase of the solar cycle and during solar minimum, and shows its weakest performance during solar maximum. The correlation of the error is substantially stronger with the sunspot number than with the ICME frequency.
    \item Our trained models generalize well to new unseen data, although absolute performance dropped on our test period due to it being selected from around the maximum of a strong solar cycle.
\end{enumerate}


\section*{Data Availability Statement}

Our machine learning dataset (preprocessed images, physical features), trained models, predictions, and all data required for reproducibility are published at \url{https://doi.org/10.5880/GFZ.OJSJ.2026.001} \cite{Collin26data}. 
The Python code for training, prediction, and evaluation is available at \url{https://github.com/DanielCollin96/proswin} \cite{Collin26code}. 
The Python code for downloading and preprocessing solar images and magnetograms is available at \url{https://github.com/DanielCollin96/solar_image_processing} \cite{Collin26code2}.

SDO solar images and magnetograms can be downloaded at \url{http://jsoc.stanford.edu/ajax/exportdata.html}. 
OMNIWeb solar wind measurements are available at \url{https://omniweb.gsfc.nasa.gov/ow.html} \cite{OMNI}, the OMNI\_M spacecraft positions at \url{https://omniweb.gsfc.nasa.gov/coho/}. 
WDC-SILSO sunspot numbers from the Royal Observatory of Belgium can be accessed at \url{https://www.sidc.be/SILSO/datafiles} \cite{SILSO_Sunspot_Number}. 
The Richardson \& Cane ICME list is available at \url{https://izw1.caltech.edu/ACE/ASC/DATA/level3/icmetable2.htm} \cite{RCdata}. 

The predictions from \citeA{Bailey21} can be accessed at \url{https://github.com/helioforecast/Papers/tree/master/Bailey2021_AmbSoWiML/results}, the ones from \citeA{Brown22} at \url{https://github.com/eddbrown/solar-swin-transformer-output-data}, and the ones from \citeA{Collin25} were published at \url{https://doi.org/10.5880/GFZ.2.7.2024.001} \cite{Collin24data}.

\section*{Conflict of Interest}

The authors declare no conflicts of interest relevant to this study.

\acknowledgments

We would like to thank Nathaniel Edward-Inatimi for kindly providing us with HUXt prediction data, which made the detailed model comparison possible. 
AIA and HMI data used here are available by courtesy of the AIA and HMI consortia.
We acknowledge use of NASA/GSFC's Space Physics Data Facility's OMNIWeb and COHOWeb service.
This work utilized high-performance computing resources made possible by funding from the Ministry of Science, Research and Culture of the State of Brandenburg (MWFK) and are operated by the IT Services and Operations unit of the GFZ Helmholtz Centre for Geosciences.
This work is supported by the Helmholtz Association’s Initiative and Networking Fund (INF) under the Helmholtz AI platform grant agreement (ID ZT-I-PF-5-1).
Daniel Collin acknowledges the support of the Helmholtz International Berlin Research School in Data Science (HEIBRiDS). 
Yuri Shprits acknowledges support from the German Research Foundation (DFG) through the projects RADIANCE (project number 547667751) and MIPT (project number 462853228).
Stefan J. Hofmeister acknowledges support from the National Science Foundation grant AGS-2229100. 


\appendix

\section{Solar Image Preprocessing}
\label{app:preprocessing}

We download $1024\times1024$ pixel images and preprocess the EUV images following procedures adapted from \citeA{Galvez19}, \citeA{Jarolim21}, and \citeA{Brown22}. 
The steps include: 
(1) updating the satellite pointing information; 
(2) correcting the image by applying the instrumental point-spread function \cite{Hofmeister24, Hofmeister25}; 
(3) registering the images such that the solar disk is centered and rotated with solar north up; 
(4) scaling the image such that the solar radius corresponds to 976 arcsec; 
(5) correcting for instrument degradation; 
(6) normalizing by the exposure time; 
(7) further downsampling to $512\times512$ pixels by summing in local blocks to emulate lower-resolution observations; 
(8) cropping a $300\times300$ pixel square with corners that are approximately at the edges of the solar disk; 
(9) resizing to $224\times224$ pixels using cubic spline interpolation, which is the input size for the image encoder. 
If an image is missing, we use the closest existing image, up to a gap of 24 hours, and rotate the Sun in this image to the missing time by applying the differential solar rotation rate. 
For the magnetograms, we apply the same pipeline except for the pointing correction, deconvolution, degradation correction, and exposure normalization steps (steps 1, 2, 5 and 6). 

Since the images have values spanning different orders of magnitude, we normalize them as suggested by \citeA{Jarolim21}. 
For the EUV images, we clip the values to a channel-specific upper threshold. The thresholds are 6457.5 for 171 Å, 7757.31 for 193 Å, and 6539.0 for 211 Å. 
Then, we rescale the values to $[0,1]$, apply an inverse hyperbolic sine function stretch ${\mathrm{asinh}(x/a)}/{\mathrm{asinh}(1/a)}$ with $a=0.005$, and rescale to $[-1,1]$. 
Magnetograms are clipped to the interval $[-100,100]$ and rescaled to $[-1,1]$. 
These steps enhance coronal-hole–specific features, reduce strong variations near active regions and ensure consistent numerical ranges across all image channels for neural network training.

\section{Distributional Regression}
\label{app:distributional_regression}

To formally introduce distributional regression, let $\mathcal{D}(Y,\theta)$, for the target variable $Y\in\mathbb{R}$, 
be the distributional model, where $\mathcal{D}$ is a probability distribution with parametric density function $p(Y|\theta)$, defined by the $K$-dimensional distributional parameter vector $\theta=(\theta_1,\ldots,\theta_K)^\top\in\Theta:=\Theta_1\times\ldots\times\Theta_K\subseteq\mathbb{R}^K$. 
We then relate each distributional parameter $\theta_k$, to possibly different subsets of the entire input vector $X\in\mathbb{R}^m$. 
For simplicity, let us write $X=x$ for all distributional parameters. 
This relation is established via parameter-specific predictors $\eta_k(x)$, such that $\theta_k\equiv\theta_k(x)=h_k(\eta_k(x))$. Here, the functions $h_k:\mathds{R}\mapsto\Theta_k$ are monotonic response functions, ensuring potential parameter space restrictions given through $\Theta_k$. 
We furthermore denote by $l(\theta;Y_{1:n})=\sum_{i=1}^n\log\lbrace p(y_i\mid\theta(x_i))\rbrace$ the log-likelihood given some observations $Y_{1:n}=(y_1,\ldots,y_n)^\top$ and associated inputs $\lbrace x_1,\ldots,x_n\rbrace$. 

In our model, we assume $Y \sim \mathrm{Lognormal}\big(\mu(x),\sigma(x)\big)$ within the distributional regression framework, where $Y$ denotes the SWS target variable, and $\mu(x)=h_\mu(\eta_\mu(x))$ and $\sigma(x)=h_\sigma(\eta_\sigma(x))$ are functions predicting the distributional parameters $\theta$, here $\mu$ and $\sigma$, based on the input~$x$.
To ensure that $\sigma(x)$ is positive for all $x\in\mathbb{R}^m$, we use $h_\sigma(x)=\exp(x)$, while $h_\mu$ is the identity. 
The predictor functions $\eta_\mu$ and $\eta_\sigma$ are learned using a neural network. 
The parameters of the neural network are obtained by minimizing the negative log-likelihood $-l(\mu,\sigma;Y_{1:n})$ of the observations. 

\section{Details on Training Process}
\label{app:training}

Before training, we standardize all physical input features by subtracting their mean and scaling them to unit variance. 
The target values are transformed by removing the location shift $\ell$ of the empirical log-normal distribution, applying a logarithm, and then standardizing, effectively changing the log-normal distributions to normal distributions. 
This allows the algorithm to fit normal distribution parameters, which is numerically more stable than using the original target variable directly in combination with a log-normal distribution. 
As mentioned in \ref{app:distributional_regression}, we apply an exponential function to the network's output $\sigma$, so that the model learns $\log(\sigma)$.

For a normal distribution, the parameters $\mu$ and $\sigma$ decouple due to their orthogonality, which enables sequential training. 
First, we only train the image encoder together with the $\mu$-head and discard the $\sigma$-head. 
Because $\mu$ represents the expected value of the normal distributions, we optimize it using a mean squared error loss function.

To improve the prediction of HSSs, we apply an exponential weighting scheme to the loss. 
For each HSS event (as defined in Section~\ref{sec:hss_cme}), the SWS values are scaled to $[0,1]$, and the weight function $w(y) = \alpha y^\gamma$ with $\alpha=5$ and $\gamma=3$ is applied. 
This shifts the training focus towards the HSS peaks and helps compensate for the tendency of machine learning models to underpredict extremes when trained on imbalanced data. 
The weighting is only applied during the first training stage.

In the second training stage, we freeze both the parameters of the image encoder and the trained $\mu$-head, and fit the $\sigma$-head alone. 
We use the negative log-likelihood of the normal distribution as the loss function with the log-transformed standardized SWS values as target variables. 
To prevent physically unrealistic uncertainty estimates, we impose an upper bound on the 99\% quantile of the corresponding log-normal distribution. 
Since HSSs rarely exceed 800 km/s \cite{Grandin19} and there is no information about CMEs in the input data, we set an upper limit of 850 km/s and clamp $\sigma$ accordingly, with the small margin above 800~km/s left to accommodate the higher speeds occasionally reached by ICMEs, which remain in the target distribution.

Additionally, we need to take into account that the $\mu$-head predicts the conditional mean of a normal distribution, but when we transform back the distributional parameters into those of a log-normal distribution, the expected value also becomes dependent on $\sigma$. 
Thus, we apply a correction to $\mu$, adjusting it such that the expected value of the log-normal distribution remains identical to the one learned during the first training stage. 
The mathematical details of these transformations and regularizations are given in \ref{app:regularization}.

We train for up to 200 epochs and evaluate the loss function after each epoch on the validation dataset. 
If the validation loss does not improve by more than $10^{-4}$ for 20 epochs, we stop training to avoid overfitting. 
The batch size and learning rate are determined by hyperparameter optimization, as described in Section~\ref{sec:hpo}.

\section{Parameter Transformations and Regularization}
\label{app:regularization}

We use a parametrization of the log-normal PDF that includes a location shift parameter $\ell$, given by
\begin{equation}\label{eq:pdf}
    p(y\mid \mu,\sigma)=\frac{1}{(y-\ell)\sigma\sqrt{2\pi}}\exp\left\{-\frac{\left(\log(y-\ell)-\mu\right)^2}{2\sigma^2}\right\} \quad \text{for } y>\ell.
\end{equation}
The parameter $\ell$ is estimated by finding the best-fitting density $p$ of on the training data. Then, the target distribution is transformed to a normal distribution by applying
\begin{equation*}
    f_\text{trans}(y)=\frac{\log(y-\ell)-\bar m}{\bar s},
\end{equation*}
to the targets, where $\bar m$ and $\bar s$ are the empirical mean and standard deviation of the log-transformed target $\log(y-\ell)$. In other words, we eliminate the location shift and standardize the targets. 

The output of the neural network, which are the mean $\tilde \mu$ and standard deviation $\tilde \sigma$ of the standardized normal distribution, can be transformed back to the actual log-normal distributional parameters $\hat\mu$ and $\hat\sigma$ for the SWS prediction by means of
\begin{equation}\label{eq:backtrafo}
    \hat\mu = \bar s \tilde \mu + \bar m,\qquad \hat\sigma=\bar s \tilde\sigma.
\end{equation}
Together with the initially estimated location parameter $\ell$, this defines the conditional distribution via the PDF.

In the following, we need to differentiate between the output $\tilde \mu$ and $\tilde \sigma$ of the neural network, which are the distributional parameters of the normal distributions we fit during the training process, and the backtransformed parameters $\hat \mu$ and $\hat \sigma$, which are the distributional parameters of the log-normal distributions we use as the actual SWS predictions, as defined in the PDF in equation (\ref{eq:pdf}).

To calculate the upper bounds on $\tilde\sigma$ during the training process, we need to find a bound that limits the 99\%-quantile $Q_{0.99}$ of the predicted log-normal distribution to the chosen physical bound $B$, i.e.,
\begin{equation*}
    Q_{0.99}[Y\mid\hat\mu,\hat\sigma]=\ell+\exp(\hat\mu+\hat\sigma z_{0.99})\overset{!}{\leq}B,
\end{equation*}
where $z_{0.99}$ is the  99\%-quantile of the standard normal distribution. 
Using equation (\ref{eq:backtrafo}) and solving for $\tilde \sigma$ yields
\begin{equation*}
    \tilde\sigma \leq \frac{\log(B-\ell)-\bar s\tilde\mu - \bar m}{\bar sz_{0.99}},
\end{equation*}
which we ensure by cropping $\tilde \sigma$ in each iteration of training and during inference. 

The $\mu$-head predicts $\tilde \mu$ as the conditional mean of a normal distribution, which yields the value $\exp(\bar s \tilde\mu+\bar m)+\ell=:E$ when scaled back to the original target distribution. We need to ensure that the expected value $\mathbb{E}[Y]$ of our predicted log-normal distribution is equal to $E$. However, the expected value of a log-normal distribution is also influenced by $\hat\sigma$. Thus, after computing $\tilde\sigma$, we correct $\tilde\mu$ such that the expected value of our log-normal prediction becomes $E$, i.e.,
\begin{equation*}
    \mathbb{E}[Y\mid\hat\mu_\text{corrected},\hat\sigma]=\ell+\exp\left(\hat\mu_\text{corrected}+\frac{\hat\sigma^2}{2}\right)\overset{!}{=}\exp(\bar s \tilde\mu_\text{old}+\bar m)+\ell.
\end{equation*}
Using again equation (\ref{eq:backtrafo}) and solving for $\tilde \mu_\text{corrected}$ yields
\begin{equation*}
    \tilde \mu_\text{corrected} =  \tilde\mu_\text{old} - \frac{\bar s \tilde\sigma^2}{2}.
\end{equation*}
The correction is applied after the cropping of $\tilde \sigma$. The combination of both steps ensures that the predicted distributions do not have unphysically large uncertainties and that the expected value remains unchanged during the training phase of the $\sigma$-head.

\section{Probabilistic Prediction}
\label{app:pred}

After training, the neural network can be used for inference of arbitrary inputs $x^\ast$, i.e., solar images and physical input features at a particular time point, and the output can be transformed back into log-normal parameter estimates $\hat\mu(x^\ast)$ and $\hat\sigma(x^\ast)$.
These yield the prediction of a conditional SWS distribution $V_{\text{sw}}\sim \mathrm{Lognormal}(\hat\mu,\hat\sigma)$, characterized by the probability density function (PDF)
\begin{equation*}
    p(y\mid\hat\mu,\hat\sigma) = \frac{1}{(y-\ell)\hat\sigma\sqrt{2\pi}}\exp\left(-\frac{\left(\log(y-\ell)-\hat\mu\right)^2}{2\hat\sigma^2}\right),
\end{equation*}
where $\ell$ is the location shift parameter derived by fitting a PDF to the empirical target distribution of the training data. 
The PDF represents the relative likelihood of $V_{\text{sw}}$ attaining a specific value $y$. 
The area under the curve over an interval gives the probability of $V_{\text{sw}}$ falling within that range. 

We can now easily compute any quantity of interest for the predicted SWS distribution. For instance, the expected value is given by
\begin{equation}\label{eq:expval}
   \mathbb{E}[V_{\text{sw}}\mid\hat\mu,\hat\sigma] = \exp\left(\hat\mu+\frac{\hat\sigma^2}{2}\right)+\ell.
\end{equation}
The cumulative distribution function (CDF), which specifies the probability that $V_{\text{sw}}$ stays below a threshold $y$, is defined as
\begin{equation*}
    F(y\mid\hat\mu,\hat\sigma) = \Phi\left(\frac{\log(y-\ell)-\hat\mu}{\hat\sigma}\right),
\end{equation*}
where $\Phi$ is the CDF of the standard normal distribution.
We can also use the CDF to derive the probability that the SWS exceeds a certain threshold, by computing $1-F(y\mid\hat\mu,\hat\sigma)$.
Finally, the quantile function, which is the inverse of the CDF, returns the threshold value $y$ for a given probability $z$, such that the probability of $V_{\text{sw}}$ staying below $y$ equals $z$. 
It is defined as
\begin{equation*}
    Q(z \mid \hat{\mu}, \hat{\sigma}) = \exp\left(\hat{\mu} + \hat{\sigma} \Phi^{-1}(z)\right) + \ell, \quad z \in (0,1),
\end{equation*}
where $\Phi^{-1}$ is the inverse of the CDF or quantile function of the standard normal distribution. We can use the quantile function to construct prediction intervals (PIs) that contain $V_{\text{sw}}$ with a specified probability level $\alpha$. The $\alpha$ PI for a prediction is given by
\begin{equation*}
    \left[Q(\alpha/2\mid\hat\mu,\hat\sigma), Q(1-\alpha/2\mid\hat\mu,\hat\sigma)\right], \quad \alpha\in(0,1).
\end{equation*}

\section{Hyperparameters}
\label{app:hyperparameters}

In Tables \ref{tab:hp_image_only} and \ref{tab:hp_full_model}, the optimized hyperparameters for all models presented in Sections \ref{sec:image-only} (image-only models) and \ref{sec:full_model} (full models) are shown, respectively. 

\begin{table}[H]
\caption{\label{tab:hp_image_only}Optimized model hyperparameters using only solar images as input, used for Table~\ref{tab:nophy}.}
\centering
\adjustbox{max width=\linewidth}{%
\begin{tabular}{lcccccccc}
\toprule
 & \multicolumn{2}{c}{Hidden layer size} & \multicolumn{2}{c}{Dropout} & \multicolumn{2}{c}{Batch size} & \multicolumn{2}{c}{Learning rate} \\
\cmidrule(l{2mm}r{2mm}){2-3}
\cmidrule(l{2mm}r{2mm}){4-5}
\cmidrule(l{2mm}r{2mm}){6-7}
\cmidrule(l{2mm}r{2mm}){8-9}
Channel(s) & $\mu$ & $\sigma$ & $\mu$ & $\sigma$ & $\mu$ & $\sigma$ & $\mu$ & $\sigma$ \\
\midrule
171         & 181                    & 182                     & 0.581             & 0.714            & 64            & 16              & 5.521$\cdot 10^{-6}$        & 2.670$\cdot 10^{-7}$       \\
193         & 159                    & 147                     & 0.659             & 0.741            & 32            & 16              & 5.659$\cdot 10^{-6}$        & 2.860$\cdot 10^{-8}$       \\
211         & 142                    & 198                     & 0.627             & 0.710            & 16            & 16              & 1.902$\cdot 10^{-7}$        & 2.402$\cdot 10^{-8}$       \\
HMI         & 176                    & 100                     & 0.563             & 0.908            & 16            & 16              & 9.905$\cdot 10^{-6}$        & 6.477$\cdot 10^{-9}$       \\
171-193     & 169                    & 57                      & 0.510             & 0.879            & 64            & 16              & 3.188$\cdot 10^{-6}$        & 5.888$\cdot 10^{-9}$       \\
171-211     & 116                    & 151                     & 0.624             & 0.742            & 64            & 16              & 3.463$\cdot 10^{-6}$        & 3.200$\cdot 10^{-7}$       \\
193-211     & 144                    & 151                     & 0.506             & 0.849            & 64            & 16              & 3.801$\cdot 10^{-7}$        & 2.858$\cdot 10^{-7}$       \\
171-193-211 & 141                    & 60                      & 0.622             & 0.724            & 16            & 16              & 3.324$\cdot 10^{-7}$        & 4.316$\cdot 10^{-7}$       \\
171-193-HMI & 177                    & 115                     & 0.603             & 0.737            & 16            & 32              & 2.745$\cdot 10^{-7}$        & 3.098$\cdot 10^{-7}$       \\
171-211-HMI & 105                    & 181                     & 0.552             & 0.848            & 32            & 16              & 3.962$\cdot 10^{-6}$        & 3.554$\cdot 10^{-7}$       \\
193-211-HMI & 191                    & 190                     & 0.592             & 0.803            & 32            & 16              & 2.975$\cdot 10^{-6}$        & 8.971$\cdot 10^{-8}$      \\
\bottomrule
\end{tabular}
}%
\end{table}

\begin{table}[H]
\caption{\label{tab:hp_full_model}Optimized model hyperparameters using all available features as input used for Table~\ref{tab:full_model}).}
\centering
\adjustbox{max width=\linewidth}{%
\begin{tabular}{lcccccccc}
\toprule
 & \multicolumn{2}{c}{Hidden layer size} & \multicolumn{2}{c}{Dropout} & \multicolumn{2}{c}{Batch size} & \multicolumn{2}{c}{Learning rate} \\
\cmidrule(l{2mm}r{2mm}){2-3}
\cmidrule(l{2mm}r{2mm}){4-5}
\cmidrule(l{2mm}r{2mm}){6-7}
\cmidrule(l{2mm}r{2mm}){8-9}
Channel(s) & $\mu$ & $\sigma$ & $\mu$ & $\sigma$ & $\mu$ & $\sigma$ & $\mu$ & $\sigma$ \\
\midrule
171         & 200                    & 135                     & 0.627             & 0.769            & 64            & 16              & 1.547$\cdot 10^{-5}$        & 2.457$\cdot 10^{-7}$       \\
193         & 105                    & 75                      & 0.607             & 0.734            & 32            & 16              & 7.116$\cdot 10^{-6}$        & 4.217$\cdot 10^{-7}$       \\
211         & 128                    & 130                     & 0.731             & 0.739            & 64            & 16              & 6.870$\cdot 10^{-7}$        & 7.976$\cdot 10^{-8}$       \\
HMI         & 166                    & 148                     & 0.685             & 0.888            & 16            & 16              & 5.631$\cdot 10^{-5}$        & 1.805$\cdot 10^{-7}$       \\
171-193     & 131                    & 121                     & 0.530             & 0.772            & 16            & 16              & 1.218$\cdot 10^{-6}$        & 1.649$\cdot 10^{-8}$       \\
171-211     & 168                    & 143                     & 0.666             & 0.704            & 32            & 16              & 2.863$\cdot 10^{-6}$        & 3.634$\cdot 10^{-7}$       \\
193-211     & 182                    & 157                     & 0.558             & 0.728            & 64            & 16              & 3.639$\cdot 10^{-7}$        & 2.791$\cdot 10^{-7}$       \\
171-193-211 & 169                    & 113                     & 0.560             & 0.922            & 32            & 64              & 9.560$\cdot 10^{-7}$        & 3.356$\cdot 10^{-7}$       \\
171-193-HMI & 138                    & 162                     & 0.575             & 0.704            & 64            & 16              & 1.017$\cdot 10^{-5}$        & 1.893$\cdot 10^{-7}$       \\
171-211-HMI & 185                    & 184                     & 0.770             & 0.702            & 32            & 16              & 5.863$\cdot 10^{-6}$        & 9.806$\cdot 10^{-8}$       \\
193-211-HMI & 182                    & 147                     & 0.559             & 0.830            & 32            & 16              & 3.480$\cdot 10^{-7}$        & 3.860$\cdot 10^{-7}$      \\
\bottomrule
\end{tabular}
}%
\end{table}


\bibliography{literature}

\end{document}